\documentclass[12pt,titlepage]{article}

\usepackage{times}

\usepackage{graphicx}
\usepackage{amsfonts}
\usepackage{mychicago}
\usepackage{genetics_manu_style}

\makeatletter
\renewcommand\@makefntext[1]{\leftskip=0em\hskip0em\@makefnmark#1}
\makeatother

\usepackage[labelfont=bf,labelsep=period,justification=raggedright,singlelinecheck=false]{caption}
\usepackage{float}

\newcommand{\singlespace}{\renewcommand{\baselinestretch}{1.3}}
\let\savedCaption=\caption
\renewcommand{\caption}[1]{\singlespace\savedCaption{\small#1}}

\usepackage{amsmath,amssymb}
\usepackage{url}

\usepackage{xspace}
\def\ROLLOFF{{\em ROLLOFF}\xspace}
\def\ALDER{{\em ALDER}\xspace}
\def\subsectioncolon#1{\subsection*{#1:}}
\def\subsubsectioncolon#1{\subsubsection*{#1:}}

\title{Inferring Admixture Histories of Human Populations Using Linkage
Disequilibrium
}

\author{
Po-Ru Loh\thanks{
Department of Mathematics and Computer Science and Artificial Intelligence Laboratory, Massachusetts Institute of Technology, Cambridge, MA 02139
\newline
$^\dagger$Broad Institute, Cambridge, MA 02142
\newline
$^\ddagger$Department of Genetics, Harvard Medical School, Boston, MA 02115
\newline
$^1$These authors contributed equally to this work.
\newline
$^2$E-mail: \texttt{bab@mit.edu}, \texttt{reich@genetics.med.harvard.edu}
}$^{,1}$,
Mark Lipson$^{\ast,1}$,
Nick Patterson$^{\dagger}$,
Priya Moorjani$^{\dagger,\ddagger}$ \and
Joseph K.\ Pickrell$^{\ddagger}$,
David Reich$^{\dagger,\ddagger,2}$,
Bonnie Berger$^{\ast,\dagger,2}$
}

\renewcommand{\CorrespondingAddress}{
\begin{tabular}{ll}
Department of Mathematics 2-373       & Department of Genetics \\              
Massachusetts Institute of Technology & Harvard Medical School \\              
77 Massachusetts Avenue               & 77 Avenue Louis Pasteur \\             
Cambridge, MA 02139                   & New Research Building, 260I \\         
                                      & Boston, MA 02115 USA \\                
Tel: (617) 253-1827                   & Tel: (617) 432-1101 \\                 
Fax: (617) 258-5429                   & Fax: (617) 432-6306 \\                 
\texttt{bab@mit.edu}                  & \texttt{reich@genetics.med.harvard.edu}
\end{tabular}
\vfill}
\renewcommand{\RunningHead}{Admixture Inference Using Weighted LD}
\renewcommand{\CorrespondingAuthor}{Bonnie Berger and David Reich}
\renewcommand{\KeyWords}{Admixture, Linkage Disequilibrium}

\begin{document}

\maketitle

\begin{abstract}
Long-range migrations and the resulting admixtures between populations have been important forces shaping human genetic diversity.  Most existing methods for detecting and reconstructing historical admixture events are based on allele frequency divergences or patterns of ancestry segments in chromosomes of admixed individuals.  An emerging new approach harnesses the exponential decay of admixture-induced linkage disequilibrium (LD) as a function of genetic distance.  Here, we comprehensively develop LD-based inference into a versatile tool for investigating admixture.  We present a new weighted LD statistic that can be used to infer mixture proportions as well as dates with fewer constraints on reference populations than previous methods.  We define an LD-based three-population test for admixture and identify scenarios in which it can detect admixture events that previous formal tests cannot.  We further show that we can uncover phylogenetic relationships among populations by comparing weighted LD curves obtained using a suite of references.  Finally, we describe several improvements to the computation and fitting of weighted LD curves that greatly increase the robustness and speed of the calculations.  We implement all of these advances in a software package, \ALDER, which we validate in simulations and apply to test for admixture among all populations from the Human Genome Diversity Project (HGDP), highlighting insights into the admixture history of Central African Pygmies, Sardinians, and Japanese.
\end{abstract}

\section*{Introduction}

Admixture between previously diverged populations has been a common
feature throughout the evolution of modern humans %
and has left significant genetic traces in contemporary populations~\cite{li2008worldwide,wall2009detecting,India,green2010draft,gravel2011demographic,pugach2011dating,draft7}. Resulting
patterns of variation can provide information about migrations,
demographic histories, and natural selection and can also be a
valuable tool for association mapping of disease genes in admixed
populations~\cite{patterson2004methods}.

Recently, a variety of methods have been developed to harness
large-scale genotype data to infer admixture events in the history of
sampled populations, as well as to estimate a range of gene flow parameters,
including ages, proportions, and sources.  Some of the most popular
approaches, such as STRUCTURE \cite{pritchard2000inference} and
principal component analysis (PCA) \cite{patterson2006population}, use
clustering algorithms to identify admixed populations as intermediates
in relation to surrogate ancestral populations.
In a somewhat similar vein, local ancestry inference methods
\cite{tang2006reconstructing,sankararaman2008estimating,price2009sensitive,lawson2012inference}
analyze chromosomes of admixed individuals with the goal of recovering continuous blocks
inherited directly from each ancestral population.  Because recombination breaks
down ancestry tracts through successive generations, the time of
admixture can be inferred from the tract length distribution
\cite{pool2009inference,pugach2011dating,gravel2012population}, with
the caveat that accurate local ancestry inference becomes difficult when tracts are short or the reference populations used are highly diverged from the true mixing populations.

A third class of methods makes use of allele frequency differentiation
among populations to deduce the presence of admixture and estimate
parameters, either with likelihood-based models
\cite{chikhi2001estimation,MLadm,sousa2009approximate,wall2009detecting,laval2010formulating,gravel2011demographic}
or with phylogenetic trees built by taking moments of the site
frequency spectrum over large sets of SNPs
\cite{India,green2010draft,draft7,pickrell2012inference,MixMapper}. For
example, $f$-statistic-based three- and four-population tests for
admixture \cite{India,green2010draft,draft7} are highly sensitive in
the proper parameter regimes and when the set of sampled populations
sufficiently represents the phylogeny.  One disadvantage of
drift-based statistics, however, is that because the rate of genetic
drift depends on population size, these methods do not allow for
inference of the time that has elapsed since admixture events.

Finally, \citeN{moorjani2011history} recently proposed a fourth approach, 
using associations between pairs of loci to make inference about
admixture, which we further develop in this article.  In general, linkage disequilibrium (LD) in a population
can be generated by selection, genetic drift, or population structure,
and it is eroded by recombination.
Within a homogeneous population, steady-state neutral LD is maintained
by the balance of drift and recombination, typically becoming
negligible in humans at distances of more than a few hundred kilobases \cite{reich2001linkage,hapmap2}.  Even
if a population is currently well-mixed, however, it can retain
longer-range \emph{admixture LD} (ALD) from admixture events in its
history involving previously separated populations.  ALD is caused by
associations between nearby loci co-inherited on an intact chromosomal
block from one of the ancestral mixing populations \cite{chakraborty1988admixture}.  Recombination
breaks down these associations, leaving a signature of the time
elapsed since admixture that can be probed by aggregating pairwise LD
measurements through an appropriate weighting scheme; the resulting
weighted LD curve (as a function of genetic distance) exhibits an
exponential decay with rate constant giving the age of admixture
\cite{moorjani2011history,draft7}.  This approach to admixture dating is
similar in spirit to strategies based on local ancestry, but
LD statistics have the advantage of a simple mathematical form that
facilitates error analysis.

In this paper, we comprehensively develop LD-based admixture inference,
extending the methodology to several novel applications that
constitute a versatile set of tools for investigating admixture.  We first
propose a cleaner functional form of the underlying weighted LD
statistic and provide a precise mathematical development of its
properties.  As an immediate result of this theory, we observe that
our new weighted LD statistic can be used to infer mixture proportions
as well as dates, extending the results of \citeN{pickrell2012genetic}.  Moreover,
such inference can still be performed (albeit with reduced power) when
data are available from only the admixed population and one surrogate
ancestral population, whereas all previous techniques require at least
two such reference populations.  As a second application, we present
an LD-based three-population test for admixture with sensitivity
complementary to the 3-population $f$-statistic test
\cite{India,draft7} and characterize the scenarios in which each is
advantageous.  We further show that phylogenetic relationships between
true mixing populations and present-day references can be inferred by
comparing weighted LD curves using weights derived from a suite of
reference populations.  Finally, we describe several improvements to
the computation and fitting of weighted LD curves: we show how to
detect confounding LD from sources other than admixture, improving the
robustness of our methods in the presence of such effects, and we present a novel
fast Fourier transform-based algorithm for weighted LD computation
that reduces typical run times from hours to seconds.  We implement
all of these advances in a software package, \ALDER (Admixture-induced
Linkage Disequilibrium for Evolutionary Relationships).

We demonstrate the performance of \ALDER by using it to test for admixture among all HGDP
populations \cite{li2008worldwide} and compare its results to those of the 3-population
test, highlighting the sensitivity trade-offs of each approach.  We
further illustrate our methodology with case studies of
Central African Pygmies,
Sardinians,
and Japanese,
revealing new details that add to our understanding of admixture events in the
history of each population.

\section*{Methods}

\subsectioncolon{Properties of weighted admixture LD}

In this section we introduce a weighted LD statistic that uses the decay
of LD to detect signals of admixture given SNP data
from an admixed population and reference populations.  This statistic
is similar to, but has an important difference from, the weighted LD
statistic used in \ROLLOFF~\cite{moorjani2011history,draft7}. The
formulation of our statistic is particularly important in
allowing us to use the amplitude (i.e., $y$-intercept) of the weighted LD curve to make inferences
about history. We begin by deriving quantitative mathematical
properties of this statistic that can be used to infer admixture
parameters.

\subsubsectioncolon{Basic model and notation}

\begin{figure}[H]
\begin{center}
\includegraphics[scale=1.5]{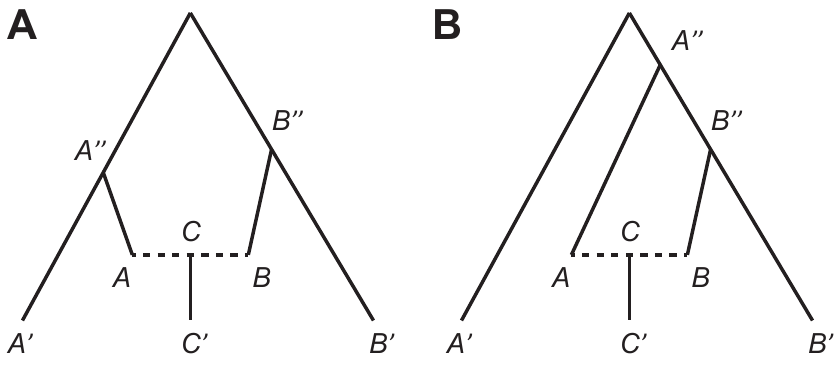}
\end{center}
\caption{Notational diagram of phylogeny containing admixed population
  and references.  Population $C'$ is descended from an
  admixture between $A$ and $B$ to form $C$; populations $A'$ and $B'$ are present-day
  references.  In practice, we assume that post-admixture drift is negligible, i.e., the $C$--$C'$ branch is extremely short and $C'$ and $C$ have identical allele frequencies.  The branch points of $A'$ and $B'$ from the $A$--$B$
  lineage are marked $A''$ and $B''$; note that in a rooted phylogeny,
  these need not be most recent common ancestors.
}
\label{fig:surrogate_ancestrals}
\end{figure}

We will primarily consider a point-admixture model in which a
population $C'$ descends from a mixture of populations $A$ and $B$ to form $C$, $n$
generations ago, in proportions $\alpha+\beta=1$, followed by random
mating (Figure~\ref{fig:surrogate_ancestrals}).  As we discuss later, we can assume for our purposes that the genetic drift between $C$ and $C'$ is negligible, and hence we will simply refer to the descendant population as $C$ as well; we will state whether we mean the population immediately after admixture vs.\ $n$ generations later when there is any risk of ambiguity.  We are interested in the properties of the LD in population
$C$ induced by admixture.  Consider two biallelic, neutrally evolving
SNPs $x$ and $y$, and for each SNP call one allele `0' and the other
`1' (this assignment is arbitrary; `0' and `1' do not need to be oriented with regard to ancestral state via an outgroup).  Denote by $p_A(x)$, $p_B(x)$, $p_A(y)$, and $p_B(y)$ the
frequencies of the `1' alleles at $x$ and $y$ in the mixing
populations $A$ and $B$ (at the time of admixture), and let $\delta(x)
:= p_A(x)-p_B(x)$ and $\delta(y) := p_A(y)-p_B(y)$ be the allele
frequency differences.

Let $d$ denote the genetic distance between $x$ and $y$ and assume
that $x$ and $y$ were in linkage equilibrium in populations $A$ and
$B$.  
Then the LD in population $C$ immediately after
admixture is
\[
D_0 = \alpha\beta\delta(x)\delta(y),
\]
where $D$ is the standard haploid measure of linkage disequilibrium as the
covariance of alleles at $x$ and $y$ \cite{chakraborty1988admixture}.
After $n$ generations of random mating, the LD decays to
\[
D_n = e^{-nd} D_0 = e^{-nd} \alpha\beta\delta(x)\delta(y)
\]
assuming infinite population size \cite{chakraborty1988admixture}.
For a finite population, the above formula holds in expectation with respect to random drift, with a
small adjustment factor caused by post-admixture drift \cite{ohta1971linkage}:
\[
E[D_n] = e^{-nd}e^{-n/2N_e}\alpha\beta\delta(x)\delta(y),
\]
where $N_e$ is the effective population size.  In most
applications the adjustment factor $e^{-n/2N_e}$ is negligible, so we
will omit it in what follows \cite[, Note S1]{Roma}.

In practice, our data consist of unphased diploid genotypes, so we
expand our notation accordingly.  Consider sampling a random
individual from population $C$ ($n$ generations after admixture).  We use a pair of $\{0, 1\}$ random
variables $X_1$ and $X_2$ to refer to the two alleles at $x$ and
define random variables $Y_1$ and $Y_2$ likewise.  Our unphased SNP
data represent observations of the $\{0,1,2\}$ random variables $X :=
X_1+X_2$ and $Y := Y_1+Y_2$.

Define $z(x,y)$ to be the covariance
\begin{equation} \label{eq:def_zxy}
z(x,y) := \text{cov}(X,Y) = \text{cov}(X_1+X_2,Y_1+Y_2),
\end{equation}
which can be decomposed into a sum of four haplotype covariances:
\begin{equation} \label{eq:diploid_cov}
z(x,y) = \text{cov}(X_1,Y_1) + \text{cov}(X_2,Y_2) + \text{cov}(X_1,Y_2) + \text{cov}(X_2,Y_1).
\end{equation}
The first two terms measure $D$ for the separate chromosomes, while
the third and fourth terms vanish, since they represent covariances
between variables for different chromosomes, which are independent.
Thus, the expectation (again with respect to random drift) of the total diploid covariance is simply
\begin{equation} \label{eq:E_zxy}
E[z(x,y)] =  2e^{-nd}\alpha\beta\delta(x)\delta(y).
\end{equation}

\subsubsectioncolon{Relating weighted LD to admixture parameters}

\citeN{moorjani2011history} first observed that pairwise LD
measurements across a panel of SNPs can be combined to enable accurate
inference of the age of admixture, $n$.  The crux of their approach
was to harness the fact that the ALD between two sites $x$ and $y$
scales as $e^{-nd}$ multiplied by the product of allele frequency
differences $\delta(x)\delta(y)$ in the mixing populations.  While the
allele frequency differences $\delta(\cdot)$ are usually not directly
computable, they can often be approximated.  Thus,
\citeN{moorjani2011history} formulated a method, \ROLLOFF, that dates
admixture by fitting an exponential decay $e^{-nd}$ to correlation
coefficients between LD measurements and surrogates for
$\delta(x)\delta(y)$.  Note that \citeN{moorjani2011history} define
$z(x,y)$ as a sample correlation coefficient, analogous to the
classical LD measure $r$, as opposed to the sample covariance
\eqref{eq:def_zxy} we use here; we find the latter more mathematically
convenient.

We build upon these previous results by deriving exact formulas for
weighted sums of ALD under a variety of weighting schemes that serve
as useful surrogates for $\delta(x)\delta(y)$ in practice.  These
calculations will allow us to interpret the magnitude of weighted ALD
to obtain additional information about admixture parameters.  Additionally,
the theoretical development will generally elucidate the behavior of
weighted ALD and its applicability in various phylogenetic scenarios.

Following \citeN{moorjani2011history}, we partition all pairs of SNPs
$(x,y)$ into bins of roughly constant genetic distance:
\[
\mathcal{S}(d) := \left\{(x,y) : d-\frac{\epsilon}{2} < |x-y| < d+\frac{\epsilon}{2}\right\},
\]
where $\epsilon$ is a discretization parameter inducing a discretization on $d$.
Given a choice of weights $w(\cdot)$, one per SNP, we define the
weighted LD at distance $d$ as
\[
a(d) := \frac{\sum_{\mathcal{S}(d)} z(x,y)w(x)w(y)}{|\mathcal{S}(d)|}.
\]

Assume first that our weights are the true allele frequency
differences in the mixing populations, i.e., $w(x) = \delta(x)$ for
all $x$.  Applying \eqref{eq:E_zxy},
\begin{eqnarray}
E[a(d)] & = & E\left[\frac{\sum_{\mathcal{S}(d)}{z(x,y)\delta(x)\delta(y)}}{|\mathcal{S}(d)|}\right] \nonumber \\
& = & \frac{\sum_{\mathcal{S}(d)}{2\alpha\beta E[\delta(x)^2\delta(y)^2] e^{-nd}}}{|\mathcal{S}(d)|} \nonumber \\
& = & 2\alpha\beta F_2(A,B)^2 e^{-nd}, \label{eq:E_ad_tworef_trueanc}
\end{eqnarray}
where $F_2(A,B)$ is the expected squared allele frequency difference
for a randomly drifting neutral allele, as defined in \citeN{India} and \citeN{draft7}.  Thus, $a(d)$ has
the form of an exponential decay as a function of $d$, with time
constant $n$ giving the date of admixture.

In practice, we must compute an empirical estimator of $a(d)$ from a
finite number of sampled genotypes.  Say we have a set of $m$ diploid
admixed samples from population $C$ indexed by $i = 1, \dots, m$, and
denote their genotypes at sites $x$ and $y$ by $x_i, y_i \in \{0, 1,
2\}$.  Also assume we have some finite number of reference individuals
from $A$ and $B$ with empirical mean allele frequences
$\hat{p}_A(\cdot)$ and $\hat{p}_B(\cdot)$.  Then our estimator is
\begin{equation} \label{eq:hat_alder_statistic}
\hat{a}(d) := \frac{\sum_{\mathcal{S}(d)}
\widehat{\text{cov}(X,Y)}(\hat{p}_A(x)-\hat{p}_B(x))(\hat{p}_A(y)-\hat{p}_B(y))}{|\mathcal{S}(d)|},
\end{equation}
where
\[
\widehat{\text{cov}(X,Y)} = \frac{1}{m-1}\sum_{i=1}^m (x_i-\overline{x})(y_i-\overline{y})
\]
is the usual unbiased sample covariance, so the expectation over the choice of samples satisfies $E[\hat{a}(d)] = a(d)$ (assuming no background LD, so the ALD in population $C$ is independent of the drift processes producing the weights).

The weighted sum $\sum_{\mathcal{S}(d)}z(x,y)w(x)w(y)$ is a natural
quantity to use for detecting ALD decay and is common to our weighted
LD statistic $\hat{a}(d)$ and previous formulations of \ROLLOFF.  Indeed,
for SNP pairs $(x,y)$ at a fixed distance $d$, we can think of
equation~\eqref{eq:E_zxy} as providing a simple linear regression
model between LD measurements $z(x,y)$ and allele frequency divergence
products $\delta(x)\delta(y)$.  In practice, the linear relation is
made noisy by random sampling, as noted above, but the regression coefficient
$2\alpha\beta e^{-nd}$ can be inferred by combining measurements from
many SNP pairs $(x,y)$.  In fact, the weighted sum
$\sum_{\mathcal{S}(d)}\hat{z}(x,y)\hat\delta(x)\hat\delta(y)$ in the numerator
 of formula \eqref{eq:hat_alder_statistic} is precisely the
numerator of the least-squares estimator of the regression
coefficient, which is the formulation of \ROLLOFF given in \citeN[,
  Note S1]{Roma}.  Note that measurements of $z(x,y)$ cannot be
combined directly without a weighting scheme, as the sign of the LD
can be either positive or negative; additionally, the weights tend to
preserve signal from ALD while depleting contributions from other
forms of LD.

Up to scaling, our \ALDER formulation is roughly equivalent to the
regression coefficient formulation of \ROLLOFF \cite[, Note S1]{Roma}.  In contrast, the
original \ROLLOFF statistic \cite{draft7} computed a {\em correlation} coefficient
between $z(x,y)$ and $w(x)w(y)$ over $\mathcal{S}(d)$.  However, the
normalization term $\sqrt{\sum_{\mathcal{S}(d)} z(x,y)^2}$ in the
denominator of the correlation coefficient can exhibit an unwanted
$d$-dependence that biases the inferred admixture date if the admixed
population has undergone a strong bottleneck \cite[, Note S1]{Roma}
or in the case of recent admixture and large sample sizes.  Beyond
correcting the date bias, the $\hat{a}(d)$ curve that \ALDER computes has
the advantage of a simple form for its amplitude in terms of
meaningful quantities, providing us additional leverage on admixture
parameters.  Additionally, we will show that $\hat{a}(d)$ can be computed
efficiently via a new fast Fourier transform-based algorithm.

\subsubsectioncolon{Using weights derived from diverged reference populations}

In the above development, we set the weights $w(x)$ to equal the
allele frequency differences $\delta(x)$ between the true mixing
populations $A$ and $B$.  In practice, in the absence of DNA samples from 
past populations, it is impossible to measure historical
allele frequencies from the time of mixture, so
instead, we substitute reference populations $A'$ and $B'$ that are accessible,
setting $w(x) = \delta'(x) := p_{A'}(x)-p_{B'}(x)$.  In a given
data set, the closest surrogates $A'$ and $B'$ may be somewhat
diverged from $A$ and $B$, so it is important to understand the
consequences for the weighted LD $a(d)$.

We show in Appendix~1 that with reference populations $A'$ and $B'$
in place of $A$ and $B$, equation~\eqref{eq:E_ad_tworef_trueanc} for
the expected weighted LD curve changes only slightly, becoming
\begin{equation} \label{eq:E_ad_tworef}
E[a(d)] = 2\alpha\beta F_2(A'',B'')^2 e^{-nd},
\end{equation}
where $A''$ and $B''$ are the branch points of $A'$ and $B'$ 
on the $A$--$B$ lineage (Figure~\ref{fig:surrogate_ancestrals}).
Notably, the curve still has the form of an exponential decay with
time constant $n$ (the age of admixture), albeit with its amplitude
(and therefore signal-to-noise ratio) attenuated according to how far $A''$
and $B''$ are from the true ancestral mixing populations.  Drift along the $A'$--$A''$
and $B'$--$B''$ branches likewise decreases signal-to-noise but in the
reverse manner: higher drift on these branches makes the weighted LD curve
noisier but does not change its expected amplitude (Figure~\ref{fig:macs_sims:amp_diverged_refs}; see Appendix~3 for additional discussion).
As above, given a real data set containing finite samples, we compute
an estimator $\hat{a}(d)$ analogous to formula \eqref{eq:hat_alder_statistic}
that has the same expectation (over sampling and drift) as the expectation of $a(d)$ with respect to drift \eqref{eq:E_ad_tworef}.

\subsubsectioncolon{Using the admixed population as one reference}

Weighted LD can also be computed with only a single reference
population by using the admixed population as the other reference
\cite[, Supplement Sec.\ 4]{pickrell2012genetic}.  Assuming first
that we know the allele frequencies of the ancestral mixing population
$A$ and the admixed population $C$, the formula for the expected curve
becomes
\begin{equation} \label{eq:E_ad_oneref_trueanc}
E[a(d)] = 2\alpha\beta^3 F_2(A,B)^2 e^{-nd}.
\end{equation}
Using $C$ itself as one reference population and $R'$ as the other
reference (which could branch anywhere between $A$ and $B$), the
formula for the amplitude is slightly more complicated, but 
the curve retains the $e^{-nd}$ decay (Figure~\ref{fig:alder_parabola}):
\begin{equation} \label{eq:E_ad_oneref}
E[a(d)] = 2\alpha\beta(\alpha F_2(A, R'') - \beta F_2(B, R''))^2 e^{-nd}.
\end{equation}
Derivations of these formulas are given in Appendix~1.

A subtle but important technical issue arises when computing
weighted LD with a single reference.  In this case, the true weighted LD statistic is
\[
a(d) = \text{cov}(X,Y)(\mu_x-p(x))(\mu_y-p(y)),
\]
where
\[
\mu_x = \alpha p_A(x)+\beta p_B(x) \quad \text{and} \quad \mu_y = \alpha p_A(y)+\beta p_B(y)
\]
are the mean allele frequencies of the admixed population (ignoring
drift) and $p(\cdot)$ denotes allele frequencies of the reference
population.  Here $a(d)$ cannot be estimated accurately by the na\"{i}ve formula
\[
\widehat{\text{cov}(X,Y)}(\hat\mu_x-\hat{p}(x))(\hat\mu_y-\hat{p}(y)),
\]
which is the natural analog of
\eqref{eq:hat_alder_statistic}.  The difficulty is that the covariance
term and the weights both involve the allele frequencies $\mu_x$ and
$\mu_y$; thus, while the standard estimators for each term are
individually unbiased, their product is a biased estimate of the
weighted LD.

\citeN{pickrell2012genetic} circumvents this problem by partitioning
the admixed samples into two groups, designating one group for use as
admixed representatives and the other as a reference population; this
method eliminates bias but reduces statistical power.  We instead
compute a polyache statistic
(File~\ref{fig:single_anc_unbiased_estimator}) that provides an
unbiased estimator $\hat{a}(d)$ of the weighted LD with maximal power.

\subsubsectioncolon{Affine term in weighted LD curve from population substructure}

Weighted LD curves computed on real populations often exhibit a
nonzero horizontal asymptote contrary to the exact exponential decay
formulas we have derived above.  Such behavior can be caused by
assortative mating resulting in subpopulations structured by ancestry percentage in violation of our model.  We show in Appendix~1
that if we instead model the admixed population as consisting of
randomly mating subpopulations with heterogeneous amounts
$\alpha$---now a random variable---of mixed ancestry, our equations
for the curves take the form
\begin{equation} \label{eq:E_ad_affine}
E[a(d)] = Me^{-nd}+K,
\end{equation}
where $M$ is a coefficient representing the contribution of admixture LD and $K$ is an additional constant produced by substructure.  Conveniently, however, the sum $M+K/2$
satisfies the same equations that the coefficient of the exponential
does in the homogeneous case: adjusting
equation~\eqref{eq:E_ad_tworef} for population substructure gives
\begin{equation} \label{eq:amp_tworef}
M+K/2 = 2\alpha\beta F_2(A'',B'')^2
\end{equation}
for two-reference weighted LD, and in the one-reference case,
modifying equation~\eqref{eq:E_ad_oneref} gives
\begin{equation} \label{eq:amp_oneref}
M+K/2 = 2\alpha\beta(\alpha F_2(A, R'') - \beta F_2(B, R''))^2.
\end{equation}
For brevity, from here on we will take the amplitude of an
exponential-plus-affine curve to mean $M+K/2$.

\subsectioncolon{Admixture inference using weighted LD}

We now describe how the theory we have developed can be used to
investigate admixture.  We detail novel techniques that use weighted
LD to infer admixture parameters, test for admixture, and learn about
phylogeny.

\subsubsectioncolon{Inferring admixture dates and fractions using one or two reference populations}

As noted above, our \ALDER formulation of weighted LD hones the
original two-reference admixture dating technique of \ROLLOFF
\cite{moorjani2011history}, correcting a possible bias \cite[, Note
  S1]{Roma}, and the one-reference technique
\cite{pickrell2012genetic}, improving statistical power.
\citeN{pickrell2012genetic} also observed that weighted LD can be used
to estimate ancestral mixing fractions.  We further develop this
application now.

The main idea is to treat our expressions for the amplitude of the
weighted LD curve as equations that can be solved for the ancestry
fractions $\alpha$ and $\beta = 1-\alpha$.  First consider
two-reference weighted LD.  Given samples from an admixed population
$C$ and reference populations $A'$ and $B'$, we compute the curve
$\hat{a}(d)$ and fit it as an exponential decay plus affine term: $\hat{a}(d)
\approx \hat{M}e^{-nd} + \hat{K}$.  Let $\hat{a}_0 := \hat{M} + \hat{K}/2$ denote
the amplitude of the curve.  Then equation~\eqref{eq:amp_tworef} gives
us a quadratic equation that we can solve to obtain an estimate
$\hat\alpha$ of the mixture fraction $\alpha$,
\[
2\hat\alpha(1-\hat\alpha)F_2(A'',B'')^2 = \hat{a}_0,
\]
assuming we can estimate $F_2(A'',B'')^2$.  Typically the branch-point
populations $A''$ and $B''$ are unavailable, but their $F_2$ distance
can be computed by means of an admixture tree
\cite{draft7,pickrell2012inference,MixMapper}.  A caveat of this
approach is that $\alpha$ and $1-\alpha$ produce the same
amplitude and cannot be distinguished by this method alone;
additionally, the inversion problem is ill-conditioned near $\alpha =
0.5$, where the derivative of the quadratic vanishes.

The situation is more complicated when using the admixed population as
one reference.  First, the amplitude relation from equation
\eqref{eq:amp_oneref} gives a quartic equation in $\hat\alpha$:
\[
2\hat\alpha(1-\hat\alpha)[\hat\alpha F_2(A, R'') - (1-\hat\alpha) F_2(B, R'')]^2 = \hat{a}_0.
\]
Second, the $F_2$ distances involved are in general not possible to
calculate by solving allele frequency moment equations~\cite{draft7,MixMapper}.
In the special case that one of the true mixing populations is
available as a reference, however---i.e., $R' =
A$---\citeN{pickrell2012genetic} demonstrated that mixture fractions
can be estimated much more easily.  From equation
\eqref{eq:E_ad_oneref_trueanc}, the expected amplitude of the curve is
$2\alpha\beta^3 F_2(A,B)^2$. On the other hand, assuming no drift in
$C$ since the admixture, allele frequencies in $C$ are given by weighted averages of allele frequencies in $A$ and $B$ with weights $\alpha$ and $\beta$; thus, the squared allele frequency differences from $A$ to $B$ and $C$ satisfy
\[
F_2(A,C) = \beta^2 F_2(A,B),
\]
and $F_2(A,C)$ is estimable directly from the sample data.  Combining
these relations, we can obtain our estimate $\hat\alpha$ by solving
the equation
\begin{equation} \label{eq:oneref_mix_prop_est}
2\hat\alpha/(1-\hat\alpha) = \hat{a}_0/F_2(A,C)^2.
\end{equation}

In practice, the true mixing population $A$ is not available for
sampling, but a closely-related population $A'$ may be.  In this case,
the value of $\hat\alpha$ given by
equation~\eqref{eq:oneref_mix_prop_est} with $A'$ in place of $A$ is a
lower bound on the true mixture fraction $\alpha$ (see Appendix~1 for theoretical development and Results for simulations exploring the tightness of the bound).  This
bounding technique is the most compelling of the above mixture
fraction inference approaches, as prior methods cannot perform such
inference with only one reference population.
In contrast, when more references are available, moment-based admixture
tree-fitting methods, for example, readily estimate mixture fractions
\cite{draft7,pickrell2012inference,MixMapper}.  In such cases we
believe that existing methods are more robust than
LD-based inference, which suffers from the degeneracy of solutions
noted above; however, the weighted LD approach can provide
confirmation based on a different genetic mechanism.

\subsubsectioncolon{Testing for admixture}

Thus far, we have taken it as given that the population $C$ of
interest is admixed and developed methods for inferring admixture
parameters by fitting weighted LD curves.  Now we consider the
question of whether weighted LD can be used to determine whether
admixture occurred in the first place.  We develop a weighted LD-based
formal test for admixture that is broadly analogous to the drift-based
3-population test \cite{India,draft7} but sensitive in different
scenarios.

A complication of interpreting weighted LD is that certain demographic
events other than admixture can also produce positive weighted LD that
decays with genetic distance, particularly in the one-reference case.
Specifically, if population $C$ has experienced a recent bottleneck or
an extended period of low population size, it may contain long-range
LD.  Furthermore, this LD typically has some correlation with allele
frequencies in $C$; consequently, using $C$ as one reference in the
weighting scheme produces a spurious weighted LD signal.

In the two-reference case, LD from reduced population size in $C$ is
generally washed out by the weighting scheme assuming the reference
populations $A'$ and $B'$ are not too genetically similar to $C$.  The
reason is that the weights $\delta(\cdot) = p_{A'}(\cdot) -
p_{B'}(\cdot)$ arise from drift between $A'$ and $B'$ that is
independent of demographic events producing LD in $C$ (beyond genetic
distances that are so short that the populations share haplotypes 
descended without recombination from their common ancestral haplotype).
Thus, observing a two-reference weighted LD decay curve is generally
good evidence that population $C$ is admixed.  There is still a
caveat, however.  If $C$ and one of the references, say $A'$, share a
recent population bottleneck, then the
bottleneck-induced LD in $C$ can be correlated to the allele
frequencies of $A'$, resulting once again in spurious weighted LD.  In
fact, the one-reference example mentioned above is the limiting case
$A'=C$ of this situation.

With these considerations in mind, we propose an LD-based
three-population test for admixture that includes a series of
pre-tests safeguarding against the pathological demographies that can
produce a non-admixture weighted LD signal.  We outline the test now;
details are in Appendix~2.  Given a population $C$ to test for
admixture and references $A'$ and $B'$, the main test is whether the
observed weighted LD $\hat{a}(d)$ using $A'$--$B'$ weights is positive and
well-fit by an exponential decay curve.  We estimate a jackknife-based
$p$-value by leaving out each chromosome in turn and re-fitting the
weighted LD as an exponential decay; the jackknife
then gives
us a standard error on the fitting parameters---namely, the amplitude
and the decay constant---that we use to measure the significance of
the observed curve.

The above procedure allows us to determine whether there is sufficient
signal in the weighted LD curve to reject the null hypothesis (under
which $\hat{a}(d)$ is random ``colored'' noise in the sense that it contains
autocorrelation).  However, in order to conclude that the curve is the
result of admixture, we must rule out the possibility that it is being
produced by demography unrelated to admixture.  We therefore apply the
following pre-test procedure.  First, we determine the distance to
which LD in $C$ is significantly correlated with LD in either $A'$ or
$B'$; to minimize signal from shared demography, we subsequently
ignore data from SNP pairs at distances smaller than this correlation
threshold.  Then, we compute one-reference weighted LD curves for
population $C$ with $A'$--$C$ and $B'$--$C$ weights and check that
the curves are well-fit as exponential decays. %
In the case that $C$ is
actually admixed between populations related to $A'$ and $B'$, the
one-reference $A'$--$C$ and $B'$--$C$ curves pick up the same
$e^{-nd}$ admixture LD decay signal.  If $C$ is not admixed but has
experienced a shared bottleneck with $A'$ (producing false-positive admixture signals from the
$A'$--$B'$ and $B'$--$C$ curves), however, the $A'$--$C$ weighting
scheme is unlikely to produce a weighted LD curve, especially when fitting beyond the LD correlation threshold.

This test procedure is intended to be conservative, so that a population $C$ identified as admixed can strongly be assumed to be so, whereas if $C$ is not identified as
admixed, we are less confident in claiming that $C$ has experienced no admixture whatsoever.  In
situations where distinguishing admixture from other demography is
particularly difficult, the test will err on the side of caution;
for example, even if $C$ is admixed, the test may fail to identify $C$ as admixed if it has also
experienced a bottleneck.  Also, if a reference $A'$ shares some of
the same admixture history as $C$ or is simply very closely related to
$C$, the pre-test will typically identify long-range correlated LD and
deem $A'$ an unsuitable reference to use for testing admixture.
The behavior of the test and pre-test criteria are explored in detail with coalescent simulations in Appendix~3.

\subsubsectioncolon{Learning about phylogeny}

Given a triple of populations $(C;A',B')$, our test can identify
admixture in the test population $C$, but what does this imply about
the relationship of populations $A'$ and $B'$ to $C$?  As with the
drift-based 3-population test, test results must be interpreted
carefully: even if $C$ is admixed, this does not necessarily mean that the
reference populations $A'$ and $B'$ are closely related to the true
mixing populations.  However, computing weighted LD curves with a
suite of different references can elucidate the phylogeny of the
populations involved, since our amplitude formulas \eqref{eq:amp_tworef} and
\eqref{eq:amp_oneref} provide information about the locations on the
phylogeny at which the references diverge from the true mixing
populations.

More precisely, in the notation of
Figure~\ref{fig:surrogate_ancestrals}, the amplitude of the
two-reference weighted LD curve is $2\alpha\beta F_2(A'',B'')^2$,
which is maximized when $A'' = A$ and $B'' = B$ and is minimized when
$A'' = B''$.  So, for example, we can fix $A'$ and compute curves for
a variety of references $B'$; the larger the resulting amplitude, the
closer the branch point $B''$ is to $B$.  In the one-reference case,
as the reference $R'$ is varied, the amplitude $2\alpha\beta(\alpha
F_2(A,R'') - \beta F_2(B,R''))^2$ traces out a parabola that starts at
$2\alpha\beta^3 F_2(A,B)^2$ when $R'' = A$, decreases to a minimum
value of 0, and increases again to $2\alpha^3\beta F_2(A,B)^2$ when
$R'' = B$ (Figure~\ref{fig:alder_parabola}).  Here, the procedure is
more qualitative because the branches $F_2(A,R'')$ and $F_2(B,R'')$
are less directly useful and the mixture proportions $\alpha$ and
$\beta$ may not be known.

\subsectioncolon{Implementation of \ALDER}

We now describe some more technical details of the \ALDER software package in which we have
implemented our weighted LD methods.

\subsubsectioncolon{Fast Fourier transform algorithm for computing weighted LD}

We developed a novel algorithm that algebraically manipulates the
weighted LD statistic into a form that can be computed using a fast
Fourier transform (FFT), dramatically speeding up the computation
(File~\ref{file:fast_weighted_ld}).  The algebraic transformation is
made possible by the simple form \eqref{eq:hat_alder_statistic} of our
weighted LD statistic along with a genetic distance discretization
procedure that is similar in spirit to
\ROLLOFF \cite{moorjani2011history} but subtly different: instead of
binning the contributions of SNP pairs $(x,y)$ by discretizing the
genetic distance $|x-y|=d$, we discretize the genetic map positions
$x$ and $y$ themselves (using a default resolution of 0.05 cM) (Figure~\ref{fig:binning}).  For
two-reference weighted LD, the resulting FFT-based algorithm that we
implemented in \ALDER has computational cost that is approximately
linear in the data size; in practice, it ran three orders of magnitude
faster than \ROLLOFF on typical data sets we analyzed.

\subsubsectioncolon{Curve-fitting}

We fit discretized weighted LD curves $\hat{a}(d)$ as $\hat{M}e^{-nd}+\hat{K}$ from
equation~\eqref{eq:E_ad_affine}, using least-squares to find best-fit
parameters.  This procedure is similar to \ROLLOFF, but \ALDER makes
two important technical advances that significantly improve the
robustness of the fitting.  First, \ALDER directly estimates the
affine term $K$ that arises from the presence of subpopulations with differing ancestry percentages by using
inter-chromosome SNP pairs that are effectively at infinite genetic
distance (Appendix~1).  The algorithmic advances we implement in
\ALDER enable efficient computation of the average weighted LD over
all pairs of SNPs on different chromosomes, giving $\hat{K}$ and,
importantly, eliminating one parameter from the exponential fitting.
In practice, we have observed that \ROLLOFF fits are sometimes
sensitive to the maximum inter-SNP distance $d$ to which the weighted
LD curve is computed and fit; \ALDER eliminates this sensitivity.

Second, because background LD is present in real populations at short
genetic distances and confounds the ALD signal (interfering with parameter estimates or
producing spurious signal entirely), it is important to fit weighted
LD curves starting only at a distance beyond which background LD is
negligible.  \ROLLOFF used a fixed threshold of $d > 0.5$ cM, but some
populations have longer-range background LD (e.g., from bottlenecks),
and moreover, if a reference population is closely related to the test
population, it can produce a spurious weighted LD signal due to recent shared demography.  \ALDER therefore estimates the extent to which the test
population shares correlated LD with the reference(s) and only fits
the weighted LD curve beyond this minimum distance as in our test for
admixture (Appendix~2).

We estimate standard errors on parameter estimates by performing a
jackknife %
over the autosomes used in the analysis, leaving out
each in turn.  Note that the weighted LD measurements from individual
pairs of SNPs that go into the computed curve $\hat{a}(d)$ are not
independent of each other; however, the contributions of different
chromosomes can reasonably be assumed to be independent.

\subsectioncolon{Data sets}

We primarily applied our weighted LD techniques to a data set of 940
individuals in 53 populations from the CEPH-Human Genome Diversity
Cell Line Panel (HGDP) \cite{rosenberg2002genetic} genotyped on an
Illumina 650K SNP array \cite{li2008worldwide}.  To study the effect
of SNP ascertainment, we also analyzed the same HGDP populations
genotyped on the Affymetrix Human Origins Array \cite{draft7}.  For
some analyses we also included HapMap Phase 3 data \cite{hapmap3}
merged either with the Illumina HGDP data set, leaving approximately
600K SNPs, or with the Indian data set of \citeN{India} including 16
Andaman Islanders (9 Onge and 7 Great Andamanese), leaving 500K SNPs.

We also constructed simulated admixed chromosomes from 112 CEU and 113
YRI phased HapMap individuals using the following procedure, described
in \citeN{moorjani2011history}.  Given desired ancestry proportions
$\alpha$ and $\beta$, the age $n$ of the point admixture, and the
number $m$ of admixed individuals to simulate, we built each admixed
chromosome as a composite of chromosomal segments from the source
populations, choosing breakpoints via a Poisson process with rate
constant $n$, and sampling blocks at random according to the specified
mixture fractions.  We stipulated that no individual haplotype could
be reused at a given locus among the $m$ simulated individuals,
preventing unnaturally long identical-by-descent segments but
effectively eliminating post-admixture genetic drift.  For the short
time scales we study (admixture occurring 200 or fewer generations
ago), this approximation has little impact.  We used this method in
order to maintain some of the complications inherent in real data.

\section*{Results}

\subsectioncolon{Simulations}

First, we demonstrate the accuracy of several forms of inference from \ALDER on simulated data.  We
generated simulated genomes for mixture fractions of 75\% YRI / 25\%
CEU and 90\% YRI / 10\% CEU and admixture dates of 10, 20, 50, 100,
and 200 generations ago.  For each mixture scenario we simulated 40
admixed individuals according to the procedure above.

We first investigated the admixture dates estimated by \ALDER using a
variety of reference populations drawn from the HGDP with varying
levels of divergence from the true mixing populations.  On the African
side, we used HGDP Yoruba (21 samples; essentially the same population as HapMap YRI) and
San (5 samples); on the European side, we used French (28 samples; very close to CEU), Han (34 samples),
and Papuan (17 samples).  We computed two-reference weighted LD curves using pairs
of references, one from each group, as well as one-reference curves
using the simulated population as one reference and each of the above
HGDP populations as the other.

For the 75\% YRI mixture, estimated dates are nearly all accurate to
within 10\% (Table~\ref{tab:sim_dates_75_25}).  The noise levels of
the fitted dates (estimated by \ALDER using the jackknife) are the
lowest for the Yoruba--French curve, as expected, followed by the
one-reference curve with French, consistent with the admixed
population being mostly Yoruba.  The situation is similar but noisier
for the 90\% YRI mixture (Table~\ref{tab:sim_dates_90_10}); in this
case, the one-reference signal is quite weak with Yoruba and
undetectable with San as the reference, due to the scaling of the amplitude (equation~\eqref{eq:amp_oneref}) with the cube of the CEU mixture fraction.

We also compared fitted amplitudes of the weighted LD curves for the
same scenarios to those predicted by formulas \eqref{eq:amp_tworef}
and \eqref{eq:amp_oneref}; the accuracy trends are similar
(Tables~\ref{tab:sim_amplitudes_75_25},~\ref{tab:sim_amplitudes_90_10}).
Finally, we tested formula \eqref{eq:oneref_mix_prop_est} for bounding
mixture proportions using one-reference weighted LD amplitudes.  We
computed lower bounds on the European ancestry fraction using French,
Russian, Sardinian, and Kalash as successively more diverged
references.  As expected, the bounds are tight for the French
reference and grow successively weaker
(Tables~\ref{tab:sim_mix_props_75_25},~\ref{tab:sim_mix_props_90_10}).
We also tried lower-bounding the African ancestry using one-reference
curves with an African reference.  In general, we expect lower bounds
computed for the major ancestry proportion to be much weaker
(Appendix~1), and indeed we find this to be the case, with the only
slightly diverged Mandenka population producing extremely weak bounds.
An added complication is that the Mandenka are an admixed population
with a small amount of  West Eurasian ancestry \cite{price2009sensitive}, which is
not accounted for in the amplitude formulas we use here.

Another notable feature of \ALDER is that, to a much greater extent than $f$-statistic methods, its inference quality improves with more samples from the admixed test population.  As a demonstration of this, we simulated a larger set of 100 admixed individuals as above, for both 75\% YRI / 25\% CEU and 90\% YRI / 10\% CEU scenarios, and compared the date estimates obtained on subsets of 5--100 of these individuals with two different reference pairs (Tables~\ref{tab:sim_vary_nadmix_dates_75_25},~\ref{tab:sim_vary_nadmix_dates_90_10}).  With larger sample sizes, the estimates become almost uniformly more accurate, with smaller standard errors.  By contrast, we observed that while using a very small sample size (say 5) for the reference populations does create noticeable noise, using 20 samples already gives allele frequency estimates accurate enough that adding more reference samples has only minimal effects on the performance of \ALDER.  This is similar to the phenomenon that the precision of $f$-statistics does not improve appreciably with more than a moderate number of samples and is due to the inherent variability in genetic drift among different loci.

\subsectioncolon{Robustness}

\begin{figure}[H]
\begin{center}
\includegraphics[width=\textwidth]{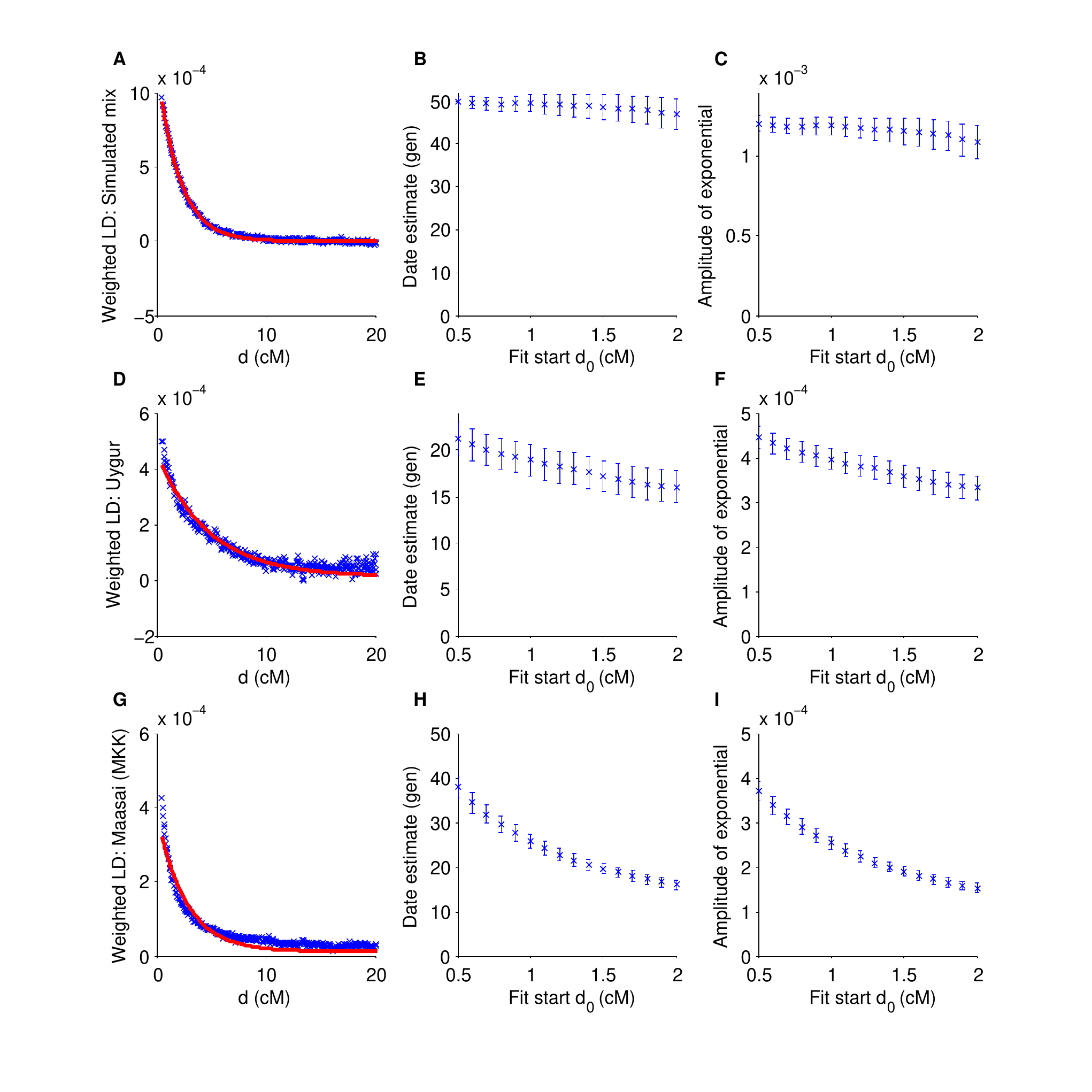}
\end{center}
\caption{Dependence of date estimates and weighted LD amplitudes on
  fitting start point.  Rows correspond to three test scenarios:
  Simulated 75\% YRI / 25\% CEU mixture 50 generations ago with
  Yoruba--French weights (top); Uygur with Han--French weights
  (middle); HapMap Maasai with Yoruba--French weights (bottom).  The
  left panel of each row shows the weighted LD curve $\hat{a}(d)$ (blue) with
  best-fit exponential decay curve (red), fit starting from $d_0 =
  0.5$ cM.  Remaining panels show the date estimate (middle) and
  amplitude (right) as a function of $d_0$. (We note that our date estimates for Uygur are somewhat more recent than those in Patterson et al.\ (2012), most likely because of our direct estimate of the affine term in the weighted LD curve.)}
\label{fig:fit_start_robustness}
\end{figure}

A challenge of weighted LD analysis is that owing to various kinds of
model violation, the parameters of the exponential fit of an observed
curve $\hat{a}(d)$ may depend on the starting distance $d_0$ from which the
curve is fit.  We therefore explored the robustness of the fitting
parameters to the choice of $d_0$ in a few scenarios
(Figure~\ref{fig:fit_start_robustness}).  First, in a simulated 75\% /
25\% YRI--CEU admixture 50 generations ago, we find that the decay
constant and amplitude are both highly robust to varying $d_0$ from
0.5 to 2.0 cM (Figure~\ref{fig:fit_start_robustness}, top).  This result is not surprising because our simulated example represents a true point
admixture with minimal background LD in the admixed population.

In practice, we expect some dependence on $d_0$ due to background LD or
longer-term admixture (either continuously over a stretch of time or
in multiple waves).  Both of these will tend to increase the weighted LD
for smaller values of $d$ relative to an exact exponential curve, so that estimates of the decay constant and
amplitude will decrease as we increase the fitting start point
$d_0$; the extent to which this effect occurs will depend on the extent of
the model violation.  We studied the $d_0$-dependence for two example
admixed populations, HGDP Uygur and HapMap Maasai (MKK).  For Uygur,
the estimated decay constants and amplitudes are fairly robust to the
start point of the fitting, varying roughly by $\pm 10$\%
(Figure~\ref{fig:fit_start_robustness}, middle).  In contrast, the
estimates for Maasai vary dramatically, decreasing by more than a
factor of 2 as $d_0$ is increased from 0.5 to 2.0 cM
(Figure~\ref{fig:fit_start_robustness}, bottom).  This behavior is
likely due to multiple-wave admixture in the genetic
history of the Maasai; indeed, it is visually evident that the
weighted LD curve for Maasai deviates from an exponential fit
(Figure~\ref{fig:fit_start_robustness}) and is in fact better-fit as a
sum of exponentials.  (See Figure~\ref{fig:macs_sims:continuous} and Appendix~3 for further simulations exploring continuous admixture.)

It is also important to consider the possibility of SNP ascertainment
bias, as in any study based on allele frequencies.  We believe that
for weighted LD, ascertainment bias could have modest effects on the amplitude, which depends on $F_2$ distances~\cite{draft7,MixMapper}, 
but will not affect the estimated date.
Running \ALDER on a suite of admixed populations in the HGDP under a
variety of ascertainment schemes suggests that admixture date
estimates are indeed quite stable to ascertainment
(Table~\ref{tab:ascertainment_effect_on_date}).
Meanwhile, the amplitudes of the LD curves can scale substantially when computed under different SNP
ascertainments, but their relative values are only different for extreme cases of African vs. non-African test populations under African vs. non-African ascertainment (Table~\ref{tab:ascertainment_effect_on_amp}; cf.\ Table~2 of \citeN{draft7}).

\subsectioncolon{Admixture test results for HGDP populations}

To compare the sensitivity of our LD-based test for admixture to the $f$-statistic-based
3-population test, we ran both \ALDER and the 3-population test on all
triples of populations in the HGDP.  Interestingly, while the tests
concur on the majority of the populations they identify as admixed,
each also identifies several populations as admixed that the other
does not (Table~\ref{tab:HGDP_LD_vs_f3}), showing that the tests have
differing sensitivity to different admixture scenarios.

\subsubsectioncolon{Admixture identified only by \ALDER}

The 3-population test loses sensitivity primarily as a result of drift
since splitting from the references' lineages.  More precisely, using
the notation of Figure~\ref{fig:surrogate_ancestrals}, the
3-population test statistic $f_3(C;A',B')$ estimates the sum of two
directly competing terms: $-\alpha\beta F_2(A'',B'')$, the negative
quantity arising from admixture that we wish to detect, and
$\alpha^2F_2(A'',A) + \beta^2F_2(B'',B) + F_2(C,C')$, a positive quantity from the
``off-tree'' drift branches.  If the latter term dominates, the
3-population test will fail to detect admixture regardless of the statistical power
available.  For example, Melanesians are only found to be admixed according to the \ALDER test; the
inability of the 3-population test to identify them as admixed is likely due to long
off-tree drift from the Papuan branch prior to admixture.  The
situation is similar for the Pygmies, for whom we do not have two close references available.

Small mixture fractions also diminish the size of the admixture term
$-\alpha\beta F_2(A,B)$ relative to the off-tree drift, and we
believe this effect along with post-admixture drift may be the reason
Sardinians are detected as admixed only by \ALDER.  In the case of the
San, who have a small 
amount of Bantu admixture \cite{pickrell2012genetic},
the small mixture
fraction may again play a role, along with the lack of a reference
population closely related to the pre-admixture San, meaning that
using existing populations incurs long off-tree drift.

\subsubsectioncolon{Admixture identified only by the 3-population test}

There are also multiple reasons why the 3-population test can
identify admixture when \ALDER does not.  For the HGDP European populations
in this category (Table~\ref{tab:HGDP_LD_vs_f3}), the 3-population test
is picking up a signal of admixture identified by \citeN{draft7} and interpreted
there as a large-scale 
admixture event in Europe involving Neolithic farmers closely related to
present-day Sardinians and an ancient northern Eurasian population. This
mixture likely began quite anciently (e.g. 7,000-9,000 years ago when
agriculture arrived in Europe \cite{bramanti2009genetic,soares2010archaeogenetics,pinhasi2012genetic}), and because admixture LD
breaks down as $e^{-nd}$, where $n$ is the age of
admixture, there is nearly no LD left for \ALDER to harness beyond the correlation threshold $d_0$.  An
additional factor that may inhibit LD-based testing is that in order
to prevent false-positive identifications of admixture, \ALDER typically eliminates reference
populations that share LD (and in particular, admixture history) with
the test population, whereas the 3-population test can use such
references.

To summarize, the \ALDER and 3-population tests both analyze a test
population for admixture using two references, but they detect signal
based on different ``genetic clocks.'' The 3-population test uses
signal from genetic drift, which can detect quite old admixture but
must overcome a counteracting contribution from post-admixture and
off-tree drift.  The LD-based test uses recombination, which is
relatively unaffected by small population size-induced long drift and
has no directly competing effect, but has limited power to detect
chronologically old admixtures because of the rapid decay of the LD
curve.  Additionally, as discussed above in the context of simulation results, the LD-based test may be better suited for large data sets, since its power is enhanced more by the availability of many samples.  The tests are thus complementary and both valuable.  (See Figure~\ref{fig:macs_sims:f3_vs_alder} and Appendix~3 for further exploration.)

\subsectioncolon{Case studies}

We now present detailed results for several human populations, all of
which \ALDER identifies as admixed but are not found by the
3-population test (Table~\ref{tab:HGDP_LD_vs_f3}).  We infer dates of
admixture and in some cases gain additional historical insights.

\subsubsectioncolon{Pygmies}

\begin{figure}[H]
\begin{center}
\includegraphics[scale=0.7]{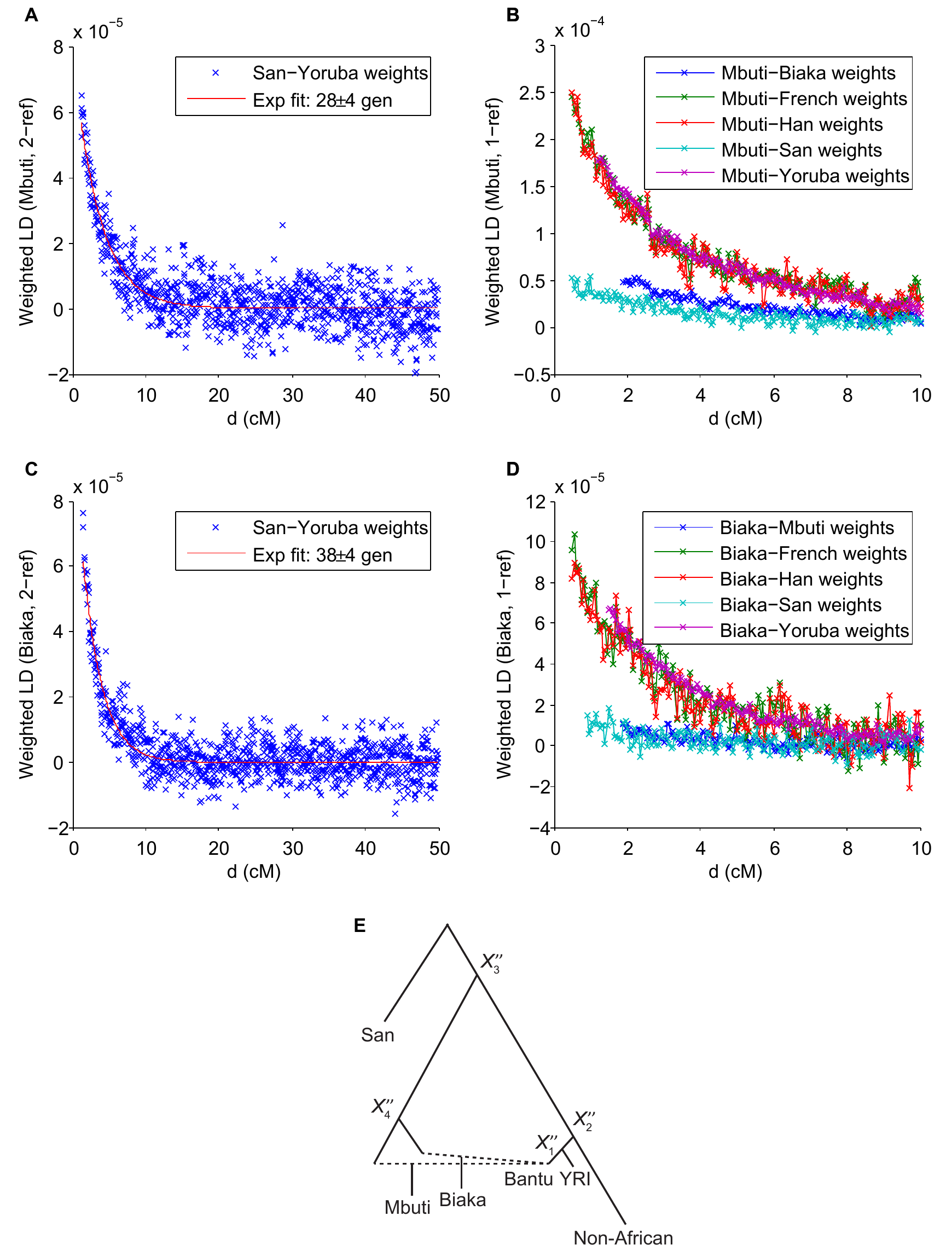}
\end{center}
\caption{Weighted LD curves for Mbuti using San and Yoruba as
  reference populations (A) and using Mbuti itself as one reference
  and several different second references (B), and analogous curves for
  Biaka (C, D). %
  Genetic distances are discretized into bins at 0.05 cM resolution.  
  Data for each curve are plotted and fit starting from the
  corresponding \ALDER-computed LD correlation thresholds.  Different
  amplitudes of one-reference curves (B, D) imply different
  phylogenetic positions of the references relative to the true mixing
  populations (i.e., different split points $X_i''$), suggesting a sketch of a putative admixture graph (E).  
  Relative branch lengths are qualitative, and the true
  root is not necessarily as depicted.}
\label{fig:Mbuti_combined_fig}
\end{figure}

Both Central African Pygmy populations in the HGDP, the Mbuti and
Biaka, show evidence of admixture (Table~\ref{tab:HGDP_LD_vs_f3}),
about $28 \pm 4$ generations (800 years) ago for Mbuti and $38 \pm 4$
generations (1100 years) ago for Biaka, estimated using San and Yoruba
as reference populations (Figure~\ref{fig:Mbuti_combined_fig}A,C).  The
intra-population heterogeneity is low, as demonstrated by the
negligible affine terms.  In each case, we also generated weighted LD
curves with the Pygmy population itself as one reference and a variety
of second references.  We found that using French, Han, or
Yoruba as the second reference gave very similar amplitudes, but the
amplitude was significantly smaller with the other Pygmy population or
San as the second reference (Figure~\ref{fig:Mbuti_combined_fig}B,D).
Using the amplitudes with Yoruba, we estimated mixture fractions of at
least $15.9\pm0.9$\% and $28.8\pm1.4$\% Yoruba-related
ancestry (lower bounds) for Mbuti and Biaka, respectively.

The phylogenetic interpretation of the relative amplitudes is
complicated by the fact that the Pygmy populations, used as
references, are themselves admixed, but a plausible coherent
explanation is as follows (see Figure~\ref{fig:Mbuti_combined_fig}E).  We
surmise that a proportion $\beta$ (bounds given above) of
Bantu-related gene flow reached the native Pygmy populations on the
order of 1000 years ago.  The common ancestors of Yoruba or
non-Africans with the Bantu population are genetically not very different from Bantu,
due to high historical population sizes (branching at positions $X_1$
and $X_2$ in Figure~\ref{fig:Mbuti_combined_fig}E).  Thus, the weighted LD
amplitudes using Yoruba or non-Africans as second references are
nearly $2\alpha^3\beta F_2(A,B)^2$, where $B$ denotes the admixing
Bantu population.  Meanwhile, San and Western (resp. Eastern) Pygmies
split from the Bantu-Mbuti (resp. Biaka) branch toward the middle or
the opposite side from Bantu ($X_3$ and $X_4$), giving a smaller
amplitude (Figure~\ref{fig:alder_parabola}).

Our results are in agreement with previous studies that have found
evidence of gene flow from agriculturalists to Pygmies
\cite{quintana2008maternal,verdu2009origins,patin2009inferring,jarvis2012patterns}.
\citeN{quintana2008maternal} suggested based on mtDNA evidence in
Mbuti that gene flow ceased several thousand years ago, but more
recently, \citeN{jarvis2012patterns} found evidence of admixture in
Western Pygmies, with a local-ancestry-inferred block length
distribution of $3.1 \pm 4.6$ Mb (mean and standard deviation),
consistent with our estimated dates.

\subsubsectioncolon{Sardinians}

\begin{figure}[H]
\begin{center}
\includegraphics{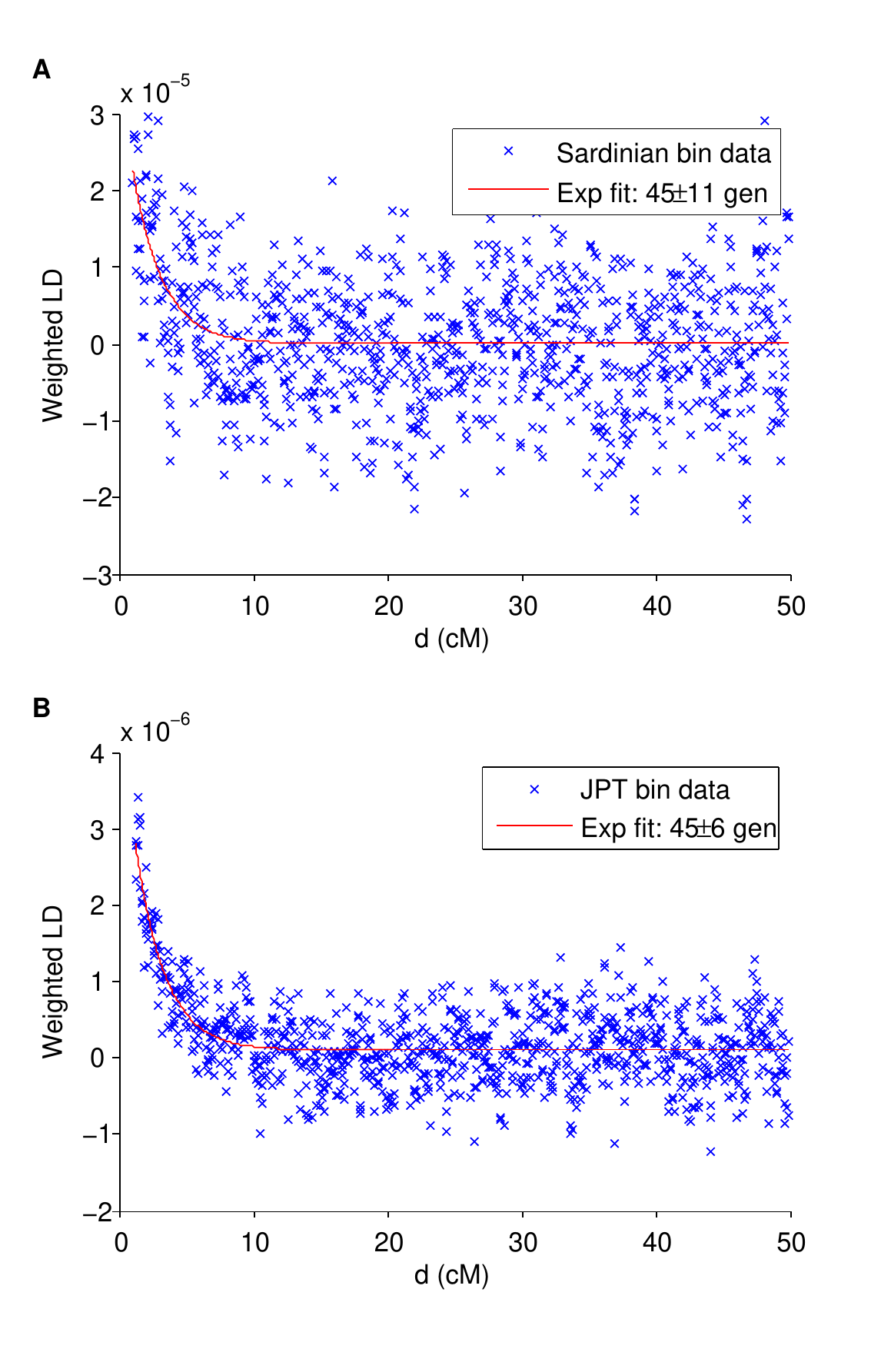}
\end{center}
\caption{Weighted LD curves for HGDP Sardinian using
  Italian--Yoruba weights (A) and HapMap Japanese (JPT) using JPT itself
  as one reference and HapMap Han Chinese (CHB) as the second reference (B).
  The exponential fits are performed starting at 1 cM and 1.2 cM, respectively, as selected by \ALDER based on detected correlated LD.}
\label{fig:Sardinian_and_JPT_curves}
\end{figure}

We detect a very small proportion of Sub-Saharan African ancestry in
Sardinians, which our \ALDER tests identified as admixed
(Table~\ref{tab:HGDP_LD_vs_f3};
Figure~\ref{fig:Sardinian_and_JPT_curves}A).  To investigate further,
we computed weighted LD curves with Sardinian as the test population and
all pairs of the HapMap CEU, YRI and CHB populations as references
(Table~\ref{tab:Sardinian_amp_tots}).  We observed an abnormally large
amount of shared long-range LD in chromosome 8, likely do to an
extended inversion segregating in Europeans \cite{price2008long}, so
we omitted it from these analyses.  The CEU--YRI curve has the largest
amplitude, suggesting both that the LD present is due to admixture and
that the small non-European ancestry component, for which we estimated
a lower bound of $0.6\pm0.2$\%, is from Africa.  (For this computation we used single-reference weighted LD with YRI as the reference, fitting the curve after 1.2 cM to reduce confounding effects from correlated LD that \ALDER detected between Sardinian and CEU.  Changing the starting point of the fit does not qualitatively affect the results.)
The existence of a
weighted LD decay curve with CHB and YRI as references provides
further evidence that the LD is not simply due to a population
bottleneck or other non-admixture sources, as does the fact that our
estimated dates from all three reference pairs are roughly consistent
at about 40 generations (1200 years) ago.  Our findings thus confirm the
signal of African ancestry in Sardinians reported in
\citeN{moorjani2011history}.  The date, small mixture proportion, and
geography are consistent with a small influx of migrants from North
Africa, who themselves traced only a fraction of their ancestry
ultimately to Sub-Saharan Africa, consistent with the findings of
\citeN{dupanloup2004estimating}.

\subsubsectioncolon{Japanese}

Genetic studies have suggested that present-day Japanese are descended
from admixture between two waves of settlers, responsible for the
Jomon and Yayoi cultures
\cite{hammer1995chromosomal,hammer2006dual,rasteiro2009revisiting}.
We also observed evidence of admixture in Japanese, and while our ability to learn about
the history was limited by the absence of a close surrogate for the
original Paleolithic mixing population, we were able to take advantage
of the one-reference inference capabilities of \ALDER.  More precisely, among our tests using all pairs of HGDP populations as references (Table~\ref{tab:HGDP_LD_vs_f3}), one reference pair, Basque and Yakut, produced a passing test for Japanese.  However, as we have noted, the reference populations need not be closely related to the true mixing populations, and we believe that in this case this seemingly odd reference pair arises as the only passing test because the data set lacks a close surrogate for Jomon.

In the absence of a reference on the Jomon side, we computed single-reference weighted LD using HapMap JPT as the test population and
JPT--CHB weights, which confer the advantage of larger sample sizes (Figure~\ref{fig:Sardinian_and_JPT_curves}B).  The weighted LD curve displays a clear decay, yielding an
estimate of $45 \pm 6$ generations, or about 1,300 years, as the age of
admixture.  To our knowledge, this is the first time genome-wide data
have been used to date admixture in Japanese. As with previous
estimates based on coalescence of Y-chromosome haplotypes
\cite{hammer2006dual}, our date is consistent with the
archaeologically attested arrival of the Yayoi in Japan roughly 2300
years ago (we suspect that our estimate is from later than the initial arrival because admixture may not have happened
immediately or may have taken place over an extended period of time). Based on the amplitude of the curve, we also obtain a (likely very conservative)
genome-wide lower bound of $41\pm3$\% ``Yayoi'' ancestry using formula
\eqref{eq:oneref_mix_prop_est} (under the reasonable assumption that
Han Chinese are fairly similar to the Yayoi population).  It is
important to note that the observation of a single-reference weighted LD
curve is not sufficient evidence to prove that a population is
admixed, but the existence of a pair of
references with which the \ALDER test identified Japanese as admixed, combined
with previous work and the lack of any signal of reduced population
size, makes us confident that our inferences are based on true
historical admixture. %

\subsubsectioncolon{Onge}

\begin{figure}[H]
\begin{center}
\includegraphics{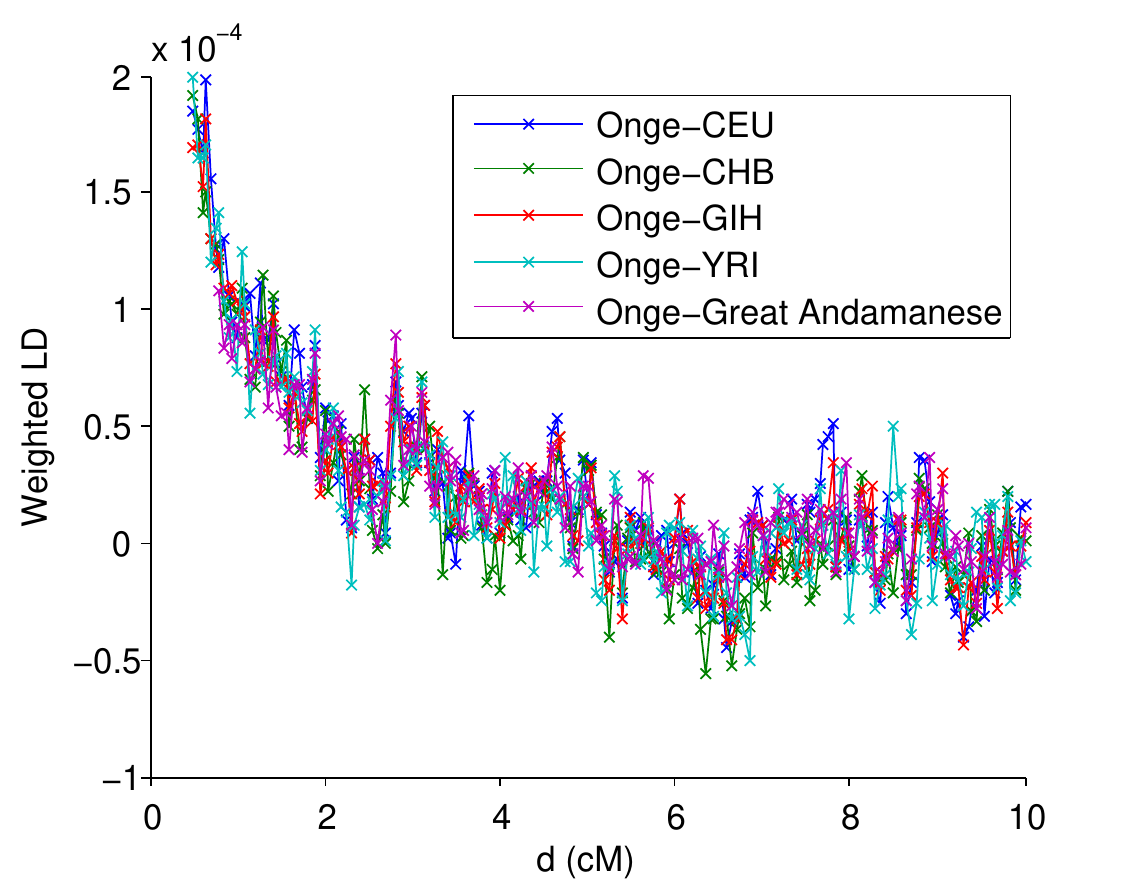}
\end{center}
\caption{Weighted LD curves for Onge using Onge itself as one
  reference and several different second references.}
\label{fig:Onge_curve}
\end{figure}

Lastly, we provide a cautionary example of weighted LD decay curves
arising from demography and not admixture.  We observed distinct
weighted LD curves when analyzing the Onge, an indigenous population
of the Andaman Islands.  However, this curve is only present when
using Onge themselves as one reference; moreover, the amplitude is
independent of whether CEU, CHB, YRI, GIH (HapMap Gujarati), or Great Andamanese is used
as the second reference (Figure~\ref{fig:Onge_curve}), as expected if
the weighted LD is due to correlation between LD and allele
frequencies in the test population alone (and independent of the
reference allele frequencies).  Correspondingly, \ALDER's LD-based test does not identify Onge as admixed using any pair of these references.  Thus, while we cannot definitively
rule out admixture, the evidence points toward internal demography
(low population size) as the cause of the elevated LD, consistent with
the current census of fewer than 100 Onge individuals.

\section*{Discussion}

\subsectioncolon{Strengths of weighted LD for admixture inference}

The statistics underlying weighted LD are quite simple, making the
formula for the expectation of $\hat{a}(d)$, as well as the noise and other
errors from our inference procedure, relatively easy to understand.
By contrast, local ancestry-based admixture dating methods (e.g.,
\citeN{pool2009inference} and \citeN{gravel2012population}) are
sensitive to imperfect ancestry inference, and it is difficult to
trace the error propagation to understand the ultimate effect on
inferred admixture parameters.
Similarly, the wavelet method of \citeN{pugach2011dating} uses
reference populations to perform (fuzzy) ancestry assignment in
windows, for which error analysis is challenging.

Another strength of our weighted LD methodology is that it has
relatively low requirements on the quality and quantity of reference
populations.  Our theory tells us exactly how the statistic behaves
for any reference populations, no matter how diverged they are from
the true ancestral mixing populations.  In contrast, the accuracy of results from
clustering and local ancestry methods is dependent on the quality of
the reference populations used in ways that are difficult to
characterize.  On the quantity side, previous approaches to admixture
inference require a surrogate for each ancestral population, whereas
as long as one is confident that the signal is truly from admixture,
weighted LD can be used with only one available reference to infer
times of admixture (as in our analysis of the Japanese) and bound
mixing fractions (as in our Pygmy case study and
\citeN{pickrell2012genetic}), problems that were previously
intractable.

Weighted LD also advances our ability to test for admixture.  As discussed above, \ALDER offers complementary sensitivity to the
3-population test and allows the identification of additional populations as admixed.  Another formal test for admixture is the 4-population test~\cite{India,draft7}, which is quite sensitive but also has trade-offs; for example, it requires three
distinctly branching references, whereas \ALDER and the 3-population
test only need two. %
Additionally, the phylogeny of the populations involved must be well understood
in order to interpret a signal of admixture from the 4-population test properly (i.e., to determine which
population is admixed).  Using weighted LD, on the other hand, largely eliminates the problem of determining the destination or direction of gene flow, since the LD signal of admixture is intrinsic to a specified test population.

\subsectioncolon{One-reference versus two-reference curves}

In practice, it is often useful to compute weighted LD curves using both the one-reference and two-reference techniques, as both can be used for inferences in different situations.  Generally, we consider two-reference curves to be more reliable for parameter estimation, since using the test population as one reference is more prone to introduce unwanted signals, such as recent admixture from a different source, non-admixture LD from reduced population size, or population structure among samples.  In particular, populations with more complicated histories and additional sources of LD beyond the specifications of our model will often have different estimates of admixture dates with one- and two-reference curves.  There is a small chance that date disagreement can reflect a false-positive admixture signal, but this is very unlikely if both one- and two-reference curves exist beyond the correlated LD threshold (see Appendix~2).  Two-reference curves also allow for direct estimation of mixture fractions, although, as discussed above, we prefer instead to use the method of single-reference bounding.

There are a number of practical considerations that make the one-reference capabilities of \ALDER desirable.  Foremost is the possibility that one may not have a good surrogate available for one of the ancestral mixing populations, as in our Japanese example.  Also, while our method of learning about phylogenetic relationships is best suited to two-reference curves because of the simpler form of the amplitude in terms of branch lengths, it is often useful to begin by computing a suite of single-reference curves, both because the data generated will scale linearly with the number of references available and because observing a range of different amplitudes gives an immediate signal of the presence of admixture in the test population.

Overall, then, a sample sequence for applying \ALDER to a new data set might be as follows: (1) test all populations for admixture using all pairs of references from among the other populations; (2) explore admixed populations of interest by comparing single-reference weighted LD curves; (3) learn more detail by analyzing selected two-reference curves alongside the one-reference ones; (4) estimate parameters using one- or two-reference curves as applicable. Of course, step (1) itself involves the complementary usefulness of both one- and two-reference weighted LD, since our test for admixture requires the presence of exponential decay signals in both types of curves.

\subsectioncolon{Effect of multiple-wave or continuous admixture}

As discussed in our section on robustness of results, in the course of our data analysis, we observed that the weighted LD
date estimate almost always becomes more recent when the exponential
decay curve is fit for a higher starting distance $d_0$.  Most
likely, this is because admixtures in human populations have taken
place over multiple generations, such that our estimated times
represent intermediate dates during the process.  To whatever extent
an admixture event is more complicated than posited in our
point-admixture model, removing low-$d$ bins will lead the fitting to
capture proportionately more of the more recent admixture.  By default, \ALDER sets $d_0$ to be the smallest distance such that non-admixture LD signals can be confidently discounted for $d > d_0$ (see Methods (Testing for admixture) and Appendix~2), but it should be noted that the selected $d_0$ will vary for different sets of populations, and in each case the true admixture signal at $d < d_0$ will also be excluded.
Theoretically, this pattern could allow us to learn more about the true
admixture history of a population, since the value of $a(d)$ at
each $d$ represents a particular function of the amount of admixture
that took place at each generation in the past.  However, in our
experience, fitting becomes difficult for any model involving more
than two or three parameters.  Thus, we made the decision to restrict
ourselves to assuming a single point admixture, fit for a principled threshold $d > d_0$, accepting that the inferred date $n$ represents
some form of average value over the true history.

\subsectioncolon{Other possible complications}

In our derivations, we have assumed implicitly that the mixing populations and the reference populations are related through a simple tree.  However, it may be that their history is more complicated, for example involving additional admixtures.  In this case, our formulas for the amplitude of the ALD curve will be inaccurate if, for example, $A$ and $A'$ have different admixture histories.  However, if our assumptions are violated only by events occurring before the divergences between the mixing populations and the corresponding references, then the amplitude will be unaffected.  Moreover, no matter what the population history, as long as $A$ and $B$ are free of measurable LD (so that our assumption of independence of alleles conditional on a single ancestry is valid), there will be no effect on the estimated date of admixture. %

\subsectioncolon{Conclusions and future directions}

In this study, we have shown how linkage disequilibrium (LD)
generated by population admixture can be a powerful tool for learning about
history, extending previous work that showed how it can be used for
estimating dates of mixture~\cite{moorjani2011history,draft7}. We have developed a new suite of
tools, implemented in the \ALDER software package, that substantially increases the speed
of admixture LD analysis, improves the robustness of admixture date inference, and exploits the
amplitude of LD as a novel source of information about history. In
particular, (a) we show how admixture LD can be leveraged into a formal test
for mixture that can sometimes find evidence of admixture not detectable by
other methods, (b) we show how to estimate mixture proportions, and (c) we
show that we can even use this information to infer phylogenetic
relationships. A limitation of \ALDER at present, however, is that it is
designed for a model of pulse admixture between two ancestral populations.
Important directions for future work will be to generalize these ideas
to make inferences about the time course of admixture in the case that it
took place over a longer period of time \cite{pool2009inference,gravel2012population} and to study multi-way admixture. In
addition, it would be valuable to be able to use the information from admixture
LD to constrain models of history for multiple populations simultaneously,
either by extending \ALDER itself or by using LD-based test results in conjunction with 
methods for fitting phylogenies incorporating admixture \cite{draft7,pickrell2012inference,MixMapper}.

\section*{Software}

Executable and C++ source files for our \ALDER software package are
available online at the Berger and Reich Lab websites: \url{http://groups.csail.mit.edu/cb/alder/}, \url{http://genetics.med.harvard.edu/reich/Reich_Lab/Software.html}.

\section*{Acknowledgments}

We are grateful to the reviewer for many suggestions that substantially increased the quality of this manuscript.
M.L.\ and P.L.\ acknowledge NSF Graduate Research Fellowship support.  P.L.\ was also partially supported by NIH training grant 5T32HG004947-04 and the Simons Foundation.  N.P., J.P.\ and
D.R.\ are grateful for support from NSF HOMINID grant \#1032255 and NIH grant
GM100233. %

\bibliographystyle{mychicago}
\bibliography{rolloff}

\section*{Appendix 1: Derivations of weighted LD formulas}

\subsectioncolon{Expected weighted LD using two diverged reference populations}

We now derive equation~\eqref{eq:E_ad_tworef} for the expected
weighted LD (with respect to random drift) using references $A'$ and $B'$ in place of $A$ and $B$,
retaining the notation of Figure~\ref{fig:surrogate_ancestrals}.
  Let $A'$ and $B'$ have allele frequencies
$p_{A'}(\cdot)$ and $p_{B'}(\cdot)$, and let $\delta'(\cdot) :=
p_{A'}(\cdot)-p_{B'}(\cdot)$ denote the allele frequency divergences
with which we weight the LD $z(x,y)$, giving the two-site statistic
\[
a(d) := z(x,y)\delta'(x)\delta'(y).
\]
(For brevity, we drop the binning procedure of averaging over SNP
pairs $(x,y)$ at distance $|x-y| \approx d$ here.)  The value of the
random variable $z(x,y)$ is affected by sampling noise as well as
genetic drift between $A$ and $B$, while the random variables
$\delta'(x)$ and $\delta'(y)$ are outcomes of genetic drift between
$A'$ and $B'$.  These random variables are uncorrelated conditional on
the allele frequencies of $x$ and $y$ in $A''$ and $B''$.  We also
assume that $x$ and $y$ are distant enough to have negligible
background LD and hence the drifts at the two sites are independent.
We then have
\begin{eqnarray*}
E[a(d)] & = & E[z(x,y)\delta'(x)\delta'(y)] \\
& = & E[E[z(x,y)\delta'(x)\delta'(y) \mid p_{A''}(x), p_{B''}(x), p_{A''}(y), p_{B''}(y)]] \\
& = & E[2\alpha\beta\delta(x)\delta(y)\delta'(x)\delta'(y)e^{-nd}] \\
& = & 2\alpha\beta e^{-nd} F_2(A'', B'')^2,
\end{eqnarray*}
where in the last step the relation $E[\delta(x)\delta'(x)] =
E[\delta(y)\delta'(y)] = F_2(A'',B'')$ follows from the fact that the
intersection of the drift paths $\delta(\cdot)$ and $\delta'(\cdot)$
is the branch between $A''$ and $B''$ \cite{India}.

\subsectioncolon{Expected weighted LD using one diverged reference population}

Using the admixed population $C$ as one reference and a population $A'$
as the other, we have $p_C(\cdot) = \alpha p_A(\cdot)+\beta
p_B(\cdot)$ (assuming negligible post-admixture drift), giving weights
\[
\delta_{A'C}(\cdot) = p_{A'}(\cdot)-\alpha p_A(\cdot)-\beta p_B(\cdot)
= \alpha \delta_{A'A}(\cdot)+ \beta \delta_{A'B}(\cdot),
\]
where $\delta_{PQ}$ denotes the allele frequency difference between
populations $P$ and $Q$.  Arguing as above, the expected weighted LD
is given by
\[
E[a(d)] = E[2\alpha\beta\delta(x)\delta(y)\delta_{A'C}(x)\delta_{A'C}(y)e^{-nd}].
\]
To complete the calculation, we compute
\[
E[\delta(\cdot)\delta_{A'C}(\cdot)]
= \alpha E[\delta(\cdot)\delta_{A'A}(\cdot)] + \beta E[\delta(\cdot)\delta_{A'B}(\cdot)].
\]
For the first term, the intersection of the $A$--$B$ and $A'$--$A$
drift paths is the $A$--$A''$ branch, so
$E[\delta(\cdot)\delta_{A'A}(\cdot)] = -F_2(A,A'')$ with the negative
sign arising because the paths traverse this branch in opposite
directions.  For the second term, the intersection of the $A$--$B$ and
$A'$--$B$ drift paths is the $A''$--$B$ branch (traversed in the same
direction), so $E[\delta(\cdot)\delta_{A'B}(\cdot)] = F_2(B,A'')$.
Combining these results gives equation~\eqref{eq:E_ad_oneref}.  (Note that a slight subtlety arises now that we are using
population $C$ in our weights: sites $x$ and $y$ can exhibit admixture
LD at appreciable distances, so $\delta_{A'C}(x)$ and
$\delta_{A'C}(y)$ are not independent.  However, only the portions of $\delta_{A'C}(x)$ and $\delta_{A'C}(y)$ arising from post-admixture drift are correlated, and this drift is negligible for typical scenarios we
study in which admixture occurred 200 or fewer generations ago.)

\subsectioncolon{Bounding mixture fractions using one reference}

We now establish our claim in the main text that the estimator
$\hat\alpha$ given in equation~\eqref{eq:oneref_mix_prop_est} for the
mixture fraction $\alpha$ is a lower bound when the reference
population $A'$ is diverged from $A$. 
Equation~\eqref{eq:oneref_mix_prop_est} gives a correct estimate when
$A' = A$ but becomes an approximation when there is genetic drift
between $A$ and $A'$ or between $C$ and $C'$. (For accuracy, in this section we relax our usual assumption of negligible drift from $C$ to $C'$.)

Rearranging equation~\eqref{eq:oneref_mix_prop_est}, we have by definition
\begin{equation} \label{eq:hatalpha_amp}
\frac{2\hat\alpha}{1-\hat\alpha} := \frac{\hat{a}_0}{F_2(A',C')^2}.
\end{equation}
From equation~\eqref{eq:E_ad_oneref_trueanc}, the amplitude
$\hat{a}_0$ is in truth given by
\[
\hat{a}_0 = 2\alpha\beta (-\alpha F_2(A,A'') + \beta F_2(B,A''))^2 e^{-n/2N_e},
\]
where we have included the post-admixture drift multiplier
$e^{-n/2N_e}$ from the $C$--$C'$ branch.  It follows that
\begin{equation} \label{eq:alpha_amp}
\frac{\hat{a}_0}{(-\alpha\beta F_2(A,A'') + \beta^2 F_2(B,A''))^2} 
= \frac{2\alpha}{\beta}e^{-n/2N_e} < \frac{2\alpha}{1-\alpha}.
\end{equation}

We claim that $F_2(A',C')^2 > (-\alpha\beta F_2(A,A'') + \beta^2
F_2(B,A''))^2$, in which case combining \eqref{eq:hatalpha_amp} and
\eqref{eq:alpha_amp} gives $\hat\alpha/(1-\hat\alpha) <
\alpha/(1-\alpha)$ and hence $\hat\alpha < \alpha$.  Indeed, we have
\begin{eqnarray*}
F_2(A',C') & > & F_2(A'',C) \\
& = & \alpha^2 F_2(A,A'') + \beta^2 F_2(B,A'') \\
& > & -\alpha\beta F_2(A,A'') + \beta^2 F_2(B,A'').
\end{eqnarray*}
Squaring both sides appears to give our claim, but we must be careful
because it is possible for the final expression to be negative.  We
will assume $A'$ is closer to $A$ than $B$, i.e., $F_2(A,A'') <
F_2(B,A'')$.  Then, if $\alpha < \beta$, the final expression is
clearly positive.  If $\alpha > \beta$, we have $\alpha^2 F_2(A,A'') >
\alpha\beta F_2(A,A'')$ and so
\[
F_2(A',C') > \alpha^2 F_2(A,A'') + \beta^2 F_2(B,A'') > \alpha\beta F_2(A,A'') - \beta^2 F_2(B,A'').
\]
Thus, squaring the inequality is valid in either case, establishing
our bound.  From the above we also see that the accuracy of the bound
depends on the sizes of the terms that are lost in the
approximation---$\alpha F_2(A,A'')$, $F_2(A',A'')$ and
$F_2(C,C')$---relative to the term that is kept, $\beta^2
F_2(B,A'')$.  In particular, aside from the bound being tighter the
closer $A'$ is to $A$, it is also more useful when the reference $A'$
comes from the minor side $\alpha < 0.5$.

\subsectioncolon{Affine term from population substructure}

In the above, we have assumed that population $C$ is homogeneously
admixed; i.e., an allele in any random admixed individual from $C$ has
a fixed probability $\alpha$ of having ancestry from $A$ and $\beta$
of having ancestry from $B$.  In practice, many admixed populations
experience assortative mating such that subgroups within the
population have varying amounts of each ancestry.  Heterogeneous
admixture among subpopulations creates LD that is independent of
genetic distance and not broken down by recombination: intuitively,
knowing the value of an allele in one individual changes the prior on
the ancestry proportions of that individual, thereby providing
information about all other alleles (even those on other chromosomes).
This phenomenon causes weighted LD curves to exhibit a nonzero
horizontal asymptote, the form of which we now derive.

We model assortative mating by taking $\alpha$ to be a random variable
rather than a fixed probability, representing the fact that
individuals from different subpopulations of $C$ have different priors
on their $A$ ancestry.  As before we set $\beta := 1-\alpha$ and we
now denote by $\bar\alpha$ and $\bar\beta$ the population-wide mean
ancestry proportions; thus, $\mu_x = \bar\alpha p_A(x)+\bar\beta
p_B(x)$.  We wish to compute the expected diploid covariance
$E[z(x,y)]$, which we saw in equation~\eqref{eq:diploid_cov} splits
into four terms corresponding to the LD between each copy of the $x$
allele and each copy of the $y$ allele.

Previously, the cross-terms $\text{cov}(X_1,Y_2)$ and
$\text{cov}(X_2,Y_1)$ vanished because a homogeneously mixed population
does not exhibit inter-chromosome LD.  Now, however, writing $\text{cov}(X_1,Y_2) = E[(X_1-\mu_x)(Y_2-\mu_y)]$ as an expectation over individuals from $C$ in the usual way, we find if we condition on the prior $\alpha$ for $A$ ancestry,
\begin{eqnarray*}
\lefteqn{E[(X_1-\mu_x)(Y_2-\mu_y) \mid p(A\text{ ancestry})=\alpha]} \\
& = & E[X_1-\mu_x \mid p(A\text{ ancestry})=\alpha] \cdot E[Y_2-\mu_y \mid p(A\text{ ancestry})=\alpha] \\
& = & (\alpha p_A(x)+\beta p_B(x)-\mu_x)(\alpha p_A(y)+\beta p_B(y)-\mu_y) \\
& = & ((\alpha-\bar\alpha)p_A(x)+(\beta-\bar\beta)p_B(x))
((\alpha-\bar\alpha)p_A(y)+(\beta-\bar\beta)p_B(y)) \\
& = & ((\alpha-\bar\alpha)p_A(x)-(\alpha-\bar\alpha)p_B(x))
((\alpha-\bar\alpha)p_A(y)-(\alpha-\bar\alpha)p_B(y)) \\
& = & (\alpha-\bar\alpha)^2\delta(x)\delta(y).
\end{eqnarray*}
That is, subpopulations with different amounts of $A$ ancestry make nonzero contributions to the covariance.  We can now compute $\text{cov}(X_1,Y_2)$ by taking the expectation of the above over the whole population (i.e., over the random variable $\alpha$):
\begin{equation} \label{eq:cov_assortative_cross_term}
\text{cov}(X_1,Y_2) = E[(\alpha-\bar\alpha)^2\delta(x)\delta(y)] = \text{var}(\alpha)\delta(x)\delta(y)
\end{equation}
and likewise for $\text{cov}(X_2,Y_1)$.

To compute the same-chromosome covariance terms, we split into two cases
according to whether or not recombination has occurred between $x$ and
$y$ since admixture.  In the case that recombination has not
occurred---i.e., the ancestry of the chromosomal region between $x$
and $y$ can be traced back as one single chunk to the time of
admixture, which occurs with probability $e^{-nd}$---the region from
$x$ to $y$ has ancestry from $A$ with probability $\alpha$ and from
$B$ with probability $\beta$.  Thus,
\begin{eqnarray*}
\lefteqn{E[(X_1-\mu_x)(Y_1-\mu_y) \mid \text{no recomb}, p(A\text{ ancestry})=\alpha]} \\
& = & \alpha E[(X_1-\mu_x)(Y_1-\mu_y) \mid A\text{ ancestry}] + \beta E[(X_1-\mu_x)(Y_1-\mu_y) \mid B\text{ ancestry}] \\
& = & \alpha (p_A(x)-\mu_x)(p_A(y)-\mu_y) + \beta (p_B(x)-\mu_x)(p_B(y)-\mu_y) \\
& = & \alpha (\bar\beta p_A(x)-\bar\beta p_B(x))(\bar\beta p_A(y)-\bar\beta p_B(y)) +
\beta (\bar\alpha p_B(x)-\bar\alpha p_a(x))(\bar\alpha p_B(y)-\bar\alpha p_A(y)) \\
& = & (\alpha\bar\beta^2+\beta\bar\alpha^2)\delta(x)\delta(y).
\end{eqnarray*}
Taking the expectation over the whole population,
\begin{equation} \label{eq:cov_assortative_no_recomb}
E[(X_1-\mu_x)(Y_1-\mu_y) \mid \text{no recomb}] = (\bar\alpha\bar\beta^2+\bar\beta\bar\alpha^2)\delta(x)\delta(y) = \bar\alpha\bar\beta\delta(x)\delta(y)
\end{equation}
as without assortative mating.

In the case where there has been a recombination, the loci are
independent conditioned upon the ancestry proportion $\alpha$, as in
our calculation of the cross-terms; hence,
\begin{equation} \label{eq:cov_assortative_recomb}
E[(X_1-\mu_x)(Y_1-\mu_y) \mid \text{recomb}] = \text{var}(\alpha)\delta(x)\delta(y),
\end{equation}
and this occurs with probability $1-e^{-nd}$.

Combining equations~\eqref{eq:cov_assortative_cross_term},
\eqref{eq:cov_assortative_no_recomb}, and
\eqref{eq:cov_assortative_recomb}, we obtain
\begin{eqnarray*}
E[z(x,y)] & = & E[(X-\mu_x)(Y-\mu_y)] \\
& = & 2\text{ var}(\alpha)\delta(x)\delta(y) + 2e^{-nd}\bar\alpha\bar\beta\delta(x)\delta(y) + 2(1-e^{-nd})\text{var}(\alpha)\delta(x)\delta(y) \\
& = & (e^{-nd}(2\bar\alpha\bar\beta-2\text{ var}(\alpha)) + 4\text{ var}(\alpha))\delta(x)\delta(y).
\end{eqnarray*}
Importantly, our final expression for $E[z(x,y)]$ still factors as the
product of a $d$-dependent term---now an exponential decay plus a
constant---and the allele frequency divergences $\delta(x)\delta(y)$.
Because it is the product $\delta(x)\delta(y)$ that interacts with our
various weighting schemes, the formulas that we have derived for the
weighted LD curve
$E[a(d)]$---equations~\eqref{eq:E_ad_tworef_trueanc},
\eqref{eq:E_ad_tworef}, \eqref{eq:E_ad_oneref_trueanc}, and
\eqref{eq:E_ad_oneref}---retain the same factors involving $F_2$
distances and change only in the replacement of $2\alpha\beta e^{-nd}$
with $e^{-nd}(2\bar\alpha\bar\beta-2\text{ var}(\alpha)) + 4\text{ var}(\alpha)$.

\section*{Appendix 2: Testing for admixture}

Here we provide details of the weighted LD-based test for
admixture we implement in \ALDER.  The test procedure is summarized in
the main text; we focus here on technical aspects not given explicitly
in Methods.

\subsectioncolon{Determining the extent of LD correlation}

The first step of \ALDER estimates the distance to which LD in the
test population is correlated with LD in each reference population.
Such correlation suggests shared demographic history that can confound
the ALD signal, so it is important to determine the distance to which
LD correlation extends and analyze weighted LD curves $\hat{a}(d)$ only for
$d$ greater than this threshold.  Our procedure is as follows.  We
successively compute LD correlation for SNP pairs $(x,y)$ within
distance bins $d_k < |x-y| < d_{k+1}$, where $d_k = kr$ for some bin
resolution $r$ (0.05 cM by default).  For each SNP pair $(x,y)$ within
a bin, we estimate the LD (i.e., sample covariance between allele
counts at $x$ and $y$) in the test population and the LD in the
reference population.  We then form the correlation coefficient
between the test LD estimates and reference LD estimates over all SNP
pairs in the bin.  We jackknife over chromosomes to estimate a
standard error on the correlation, and we set our threshold after the
second bin for which the correlation is insignificant ($p>0.05$).  To reduce dependence on sample size, we then repeat this procedure with successively increasing resolutions up to $0.1$ cM and set the final threshold as the maximum of the cutoffs obtained.

\subsectioncolon{Determining significance of a weighted LD curve}

To define a formal test for admixture based on weighted LD, we need to
estimate the significance of an observed weighted LD curve $\hat{a}(d)$.
This question is statistically subtle for several reasons.  First, the
null distribution of the curve $\hat{a}(d)$ is complex.  Clearly the test
population $C$ should not be admixed under the null hypothesis, but as
we have discussed, shared demography---particularly bottlenecks---can
also produce weighted LD.  We circumvent this issue by using the
pre-tests described in the next section and assume that if the test
triple $(C;A',B')$ passes the pre-tests, then under the null
hypothesis, non-admixture demographic events have negligible effect on
weighted LD beyond the correlation threshold computed above.  Even so,
the $\hat{a}(d)$ curve still cannot be modeled as random white noise:
because SNPs contribute to multiple bins, the curve typically exhibits
noticeable autocorrelation.  Finally, even if we ignore the issue of
colored noise, the question of distinguishing a curve of any type---in
our case, an exponential decay---from noise is technically subtle: the
difficulty is that a singularity arises in the likelihood surface
when the amplitude vanishes, which is precisely the hypothesis that we
wish to test~\cite{davies1977hypothesis}.

In light of these considerations, we estimate a $p$-value using the
following procedure, which we feel is well-justified despite not being
entirely theoretically rigorous.  We perform jackknife replicates of
the $\hat{a}(d)$ curve computation and fitting, leaving out one chromosome
in each replicate, and estimate a standard error for the amplitude and
decay constant of the curve using the usual jackknife procedure.  We
obtain a ``$z$-score'' for the amplitude and the decay constant by
dividing each by its estimated standard error.  Finally, we take
the minimum (i.e., less-significant) of these $z$-scores and convert it to a
$p$-value assuming it comes from a standard normal; we report this
$p$-value as our final significance estimate.

Our intuition for this procedure is that checking the ``$z$-score'' of
the decay constant essentially tells us whether or not the exponential
decay is well-determined: if the $\hat{a}(d)$ curve is actually just noise,
then the fitting of jackknife replicates should fluctuate
substantially.  On the other hand, if the $\hat{a}(d)$ curve has a stable
exponential decay constant, then we have good evidence that $\hat{a}(d)$ is
actually well-fit by an exponential---and in particular, the amplitude
of the exponential is nonzero, meaning we are away from the
singularity.  In this case the technical difficulty is no longer an
issue and the jackknife estimate of the amplitude should in fact give
us a good estimate of a $z$-score that is approximately normal under
the null.  The ``$z$-score'' for the decay constant certainly is not
normally distributed---in particular, it is always positive---but
taking the minimum of these two scores only makes the test more
conservative.

Perhaps most importantly, we have compelling empirical evidence that
our $z$-scores are well-behaved under the null.  We applied our test
to nine HGDP populations that neither \ALDER nor the 3-population test
identified as admixed; for each test population, we used as references
all populations with correlated LD detectable to no more than 0.5 cM.
These test triples thus comprise a suite of approximately null tests.
We computed Q-Q plots for the reported $z$-scores and observed that
for $z>0$ (our region of interest), our reported z-scores follow the
normal distribution reasonably well, generally erring slightly on the
conservative side (Figure~\ref{fig:QQ_plots}).  These findings give
strong evidence that our significance calculation is sufficiently
accurate for practical purposes; in reality, model violation is likely
to exert stronger effects than the approximation error in our
$p$-values, and although our empirical tests cannot probe the tail
behavior of our statistic, for practical purposes the precise values
of $p$-values less than, say, $10^{-6}$ are generally inconsequential.

\subsectioncolon{Pre-test thresholds}

To ensure that our test is applicable to a given triple $(C;A',B')$, we need to
rule out the possibility of demography producing
non-admixture-related weighted LD.  We do so by computing
weighted LD curves for $C$ with weights $A'$--$B'$, $A'$--$C$, and $B'$--$C$
and fitting an exponential to each curve.  To eliminate the
possibility of a shared ancestral bottleneck between $C$ and one of
the references, we check that the three estimated
amplitudes and decay constants are well-determined; explicitly, we
compute a jackknife-based standard error for each parameter and
require the implied $p$-value for the parameter being positive to be
less than 0.05. If so, we conclude that whatever LD
is present is due to admixture, not other demography, and we report
the $p$-value estimate defined above for the significance of the
$A'$--$B'$ curve as the $p$-value of our test.

We are aware of one demographic scenario in which the \ALDER test could potentially return a finding of admixture when the test population is not in fact admixed.  As illustrated in Figure~\ref{fig:pathological_demography}, this would occur when $A'$ and $C$ have experienced a shared bottleneck and $C$ has subsequently had a further period of low population size.  We do not believe that we have ever encountered such a false positive admixture signal, but to guard against it, we note that if it were to occur, the three decay time constants for the reference pairs $A'$--$B'$, $A'$--$C$, and $B'$--$C$ would disagree. Thus, along with the test results, \ALDER returns a warning whenever the three best-fit values of the decay constant do not agree to within 25\%. 

\subsectioncolon{Multiple-hypothesis correction}

In determining statistical significance of test results when testing a
population using many pairs of references, we apply a
multiple-hypothesis correction that takes into account the number of
tests being run.  Because some populations in the reference set may be
very similar, however, the tests may not be independent.  We therefore
compute an effective number $n_r$ of distinct references by running PCA on
the allele frequency matrix of the reference populations; we take $n_r$ to be the number of singular values
required to account for 90\% of the total variance. Finally, we apply a Bonferroni correction to the $p$-values from each test using the effective number $\binom{n_r}{2}$ of reference pairs.

\section*{Appendix 3: Coalescent simulations}

Here we further validate and explore the properties of weighted LD
with entirely {\em in silico} simulations using the Markovian
coalescent simulator MaCS \cite{chen2009fast}.  These simulations
complement the exposition in the main text in which we constructed
simulated admixed chromosomes by piecing together haplotype fragments
from real HapMap individuals.

\subsectioncolon{Effect of divergence and drift on weighted LD amplitude}

To illustrate the effect of using reference populations with varying
evolutionary distances from true mixing populations, we performed a set
of four simulations in which we varied one reference population in a
pair of dimensions: (1) time depth of divergence from the true
ancestor, and (2) drift since divergence.  In each case, we simulated
individuals from three populations $A'$, $C'$, and $B'$, with $22\%$
of $C'$s ancestry derived from a pulse of admixture 40 generations ago
from $B$, where $A'$ and $B'$ diverged 1000 generations ago.  We
simulated 5 chromosomes of 100 Mb each for 20 diploid individuals from
each of $A'$ and $B'$ and 30 individuals from $C'$, with diploid
genotypes produced by randomly combining pairs of haploid chromosomes.
We assumed an effective population size of 10,000 and set the
recombination rate to $10^{-8}$.  We set the mutation rate parameter
to $10^{-9}$ to have the same effect as using a mutation rate of
$10^{-8}$ and then thinning the data by a factor of 10 (as it would
otherwise have produced an unnecessarily large number of SNPs).
Finally, we set the MaCS history parameter (the Markovian order of the simulation, i.e., the distance to which the full ancestral recombination graph is maintained) to $10^4$ bases.

For the first simulation
(Figure~\ref{fig:macs_sims:amp_diverged_refs}A), we set the divergence
of $A'$ and $C'$ to be immediately prior to the gene flow event,
altogether resulting in the following MaCS command:
\begin{quote}
\footnotesize
\texttt{macs 140 1e8 -i 5 -h 1e4 -t 0.00004 -r 0.0004 -I 3 40 40 60 -em 0.001 3 2 10000 -em 0.001025 3 2 0 -ej 0.001025 1 3 -ej 0.025 2 3}
\end{quote}
For the second simulation
(Figure~\ref{fig:macs_sims:amp_diverged_refs}B), we increased the
drift along the $A'$ terminal branch by reducing the population size
by a factor of 20 for the past 40 generations:
\begin{quote}
\footnotesize
\texttt{-en 0 1 0.05 -en 0.001 1 1}
\end{quote}
For the third and fourth simulations
(Figure~\ref{fig:macs_sims:amp_diverged_refs}C,D), we changed the
divergence time of $A'$ and $C'$ from 41 to 520 generations, half the
distance to the root:
\begin{quote}
\footnotesize
\texttt{-ej 0.001025 1 3 \quad -> \quad -ej 0.013 1 3}
\end{quote}

We computed weighted LD curves using $A'$--$B'$ references
(Figure~\ref{fig:macs_sims:amp_diverged_refs}), and the results
corroborate our derivation and discussion of
equation~\eqref{eq:E_ad_tworef}.  In all cases, the estimated date of
admixture is within statistical error of the simulated 40-generation
age.  The amplitude of the weighted LD curve is unaffected by drift in
$A'$ but is substantially reduced by the shorter distance
$F_2(A'',B'')$ in the latter two simulations.  Increased drift to $A'$
does, however, make the weighted LD curves in the right two panels
somewhat noisier than the left two.

\subsectioncolon{Validation of pre-test criteria in test for admixture}

To understand the effects of the pre-test criteria stipulated in our
LD-based test for admixture, we simulated a variety of population
histories with and without mixture.  In each case we used the same
basic parameter settings as above, except we set the root of each tree
to be 4000 generations ago and we simulated 10 chromosomes for each
individual instead of 5.

\subsubsectioncolon{Scenario 1} True admixture 40 generations ago;
reference $A'$ diverged 400 generations ago (similar to
Figure~\ref{fig:macs_sims:amp_diverged_refs}C).  All pre-tests pass
and the our test correctly identifies admixture.

\subsubsectioncolon{Scenario 2} True admixture 40 generations ago;
reference $A'$ diverged 41 generations ago (similar to
Figure~\ref{fig:macs_sims:amp_diverged_refs}A).  Because of the
proximity of the admixed population $C'$ and the reference $A'$, the
test detects long-range correlated LD and concludes that using $A'$ as
a reference may produce unreliable results.

\subsubsectioncolon{Scenario 3} True admixture 40 generations ago;
contemporaneous gene flow (of half the magnitude) to the lineage of
the reference population $A'$ as well.  Again, the pre-test detects
long-range correlated LD and concludes that $A'$ is an unsuitable
reference.

\subsubsectioncolon{Scenario 4} No admixture; $A$ and $C$ simply
form a clade diverging at half the distance to the root (similar to
Figure~\ref{fig:macs_sims:amp_diverged_refs}C without the gene flow).
The test finds no evidence for admixture; weighted LD measurements do
not exhibit a decay curve.

\subsubsectioncolon{Scenario 5} No admixture; $A$ and $C$ diverged
40 generations ago.  As above, the test finds no decay in weighted LD.
In this scenario the pre-test does detect substantial correlated LD to
$1.95$ cM because of the proximity of $A$ and $C$.

\subsubsectioncolon{Scenario 6} No admixture; same setup as Scenario 4
with addition of recent bottleneck in population $C$ (100-fold reduced
population size for the past 40 generations).  Here, the test finds no
weighted LD decay in the two-reference curve and concludes that there
is no evidence for admixture.  It does, however, detect decay curves
in both one-reference curves (with $A$--$C$ and $B$--$C$ weights);
these arise because of the strong bottleneck-induced LD within
population $C$.

\subsubsectioncolon{Scenario 7} No admixture; shared bottleneck: $A$
and $C$ diverged 40 generations ago and their common ancestor
underwent a bottleneck of 100-fold reduced population size for the
preceding 40 generations.  In this case the pre-test detects an
enormous amount of correlated LD between $A$ and $C$ and deems $A$ an
unsuitable reference.

\subsectioncolon{Sensitivity comparison of 3-population test and LD-based test for admixture}

Here we compare the sensitivities of the allele frequency moment-based
3-population test \cite{India,draft7} and our LD-based test for
admixture.  We simulated a total of 450 admixture scenarios in which
we varied three parameters: the age of the branch point $A''$ (1000,
2000, and 3000 generations), the date $n$ of gene flow (20 to 300 in
increments of 20), and the fraction $\alpha$ of $A$ ancestry (50\% to
95\% in increments of 5\%), as depicted in
Figure~\ref{fig:macs_sims:f3_vs_alder}.  In each case we simulated 40
admixed individuals, otherwise using the same parameter settings as in
the scenarios above.  Explicitly, we ran the commands:
\begin{quote}
\footnotesize
\texttt{macs 160 1e8 -i 10 -h 1e4 -t 0.00004 -r 0.0004 -I 3 40 40 80 -em tMix 3 2 migRate -em tMixStop 3 2 0 -ej tSplit 1 3 -ej 0.1 2 3}
\end{quote}
where \texttt{tMix} and \texttt{tSplit} correspond to $n$ and the age
of $A''$, while \texttt{migRate} and \texttt{tMixStop} produce a pulse
of gene flow from the $B'$ branch giving $C'$ a fraction $\alpha$ of
$A$ ancestry.

We then ran both the 3-population test ($f_3$) and the \ALDER test on
$C'$ using $A'$ and $B'$ as references
(Figure~\ref{fig:macs_sims:f3_vs_alder}).  The results of these
simulations show clearly that the two tests do indeed have
complementary parameter ranges of sensitivity.  We first observe that
the $f_3$ test is essentially unaffected by the age of admixture (up
to the 300 generations we investigate here).  As discussed in the main
text, its sensitivity is constrained by competition between the
admixture signal of magnitude $\alpha\beta F_2(A'',B'')$ and the
``off-tree drift'' arising from branches off the lineage connecting
$A'$ and $B'$ \cite{India}---in this case, essentially the quantity
$\alpha^2F_2(A'',C')$.  Thus, as the divergence point $A''$ moves up
the lineage, the threshold value of $\alpha$ below which the $f_3$
test can detect mixture decreases.

The \ALDER tests behave rather differently, exhibiting a drop-off in
sensitivity as the age of admixture increases, with visible noise near
the thresholds of sufficient sensitivity.  The difference between the
$f_3$ and \ALDER results is most notable in the bottom panels of
Figure~\ref{fig:macs_sims:f3_vs_alder}, where the reference $A'$ is
substantially diverged from $C'$.  In this case, \ALDER is still able
to identify small amounts of admixture from the $B'$ branch, whereas
the $f_3$ test cannot.  Also notable are the vertical swaths of failed
tests centered near $\alpha = 0.9, 0.75$, and $0.65$ for $A''$
respectively located at distances $0.75$, $0.5$, and $0.25$ along the
branch from the root to $A'$.  This feature of the results arises
because the amplitude of the single-reference weighted LD curve with
$A'$--$C'$ weights vanishes near those values of $\alpha$ (see
equation \eqref{eq:E_ad_oneref} and Figure~\ref{fig:alder_parabola}),
causing the \ALDER pre-test to fail.  (The two-reference weighted LD
exhibits a clear decay curve, but the pre-test is being overly
conservative in these cases.)  Finally, we also observe that for the
smallest choice of mixture age (20 generations), many \ALDER tests
fail.  In these cases, the pre-test detects long-range correlated LD
with the reference $B'$ and is again overly conservative.
 
\subsectioncolon{Effect of protracted admixture on weighted LD}

The admixture model that we analyze in this manuscript treats
admixture as occurring instantaneously in a single pulse of gene flow;
however, in real human populations, admixture typically occurs
continuously over an extended period of time.  Here we explore the
effect of protracted admixture on weighted LD curves by simulating
scenarios involving continuous migration.  We used a setup nearly
identical to the simulations above for comparing the $f_3$ and \ALDER
tests, except here we modified the migration rate and start and end
times to correspond to $40\%$ $B$ ancestry that continuously mixed
into population $C$ over a period of 0--200 generations ending 40
generations ago.  We varied the duration of admixture in increments of
20 generations.

For each simulation, we used \ALDER to compute the two-reference
weighted LD curve and fit an exponential decay.  In each case the date
of admixture estimated by \ALDER (Figure~\ref{fig:macs_sims:continuous}A) falls within the time interval of
continuous mixture, as expected \cite{moorjani2011history}.  For shorter durations of
admixture spanning up to 50 generations or so, the estimated date
falls very near the middle of the interval, while it is downward
biased for mixtures extending back to hundreds of generations.  The
amplitude of the fitted exponential also exhibits a downward bias as
the mixture duration increases
(Figure~\ref{fig:macs_sims:continuous}B).  This behavior occurs
because unlike the point admixture case, in which the weighted LD
curve follows a simple exponential decay
(Figure~\ref{fig:macs_sims:continuous}C), continuous admixture creates
weighted LD that is an average of exponentials with different decay
constants (Figure~\ref{fig:macs_sims:continuous}D).

\newpage
\section*{Figures}

\newpage

\newpage

\newpage

\newpage

\newpage
\section*{Tables}

\begin{table}[H]
\caption{\bf{Results of \ALDER and 3-population tests for admixture on HGDP populations.\newline}}
{\footnotesize\sffamily
\begin{tabular}{|lrr|lr|lr|l|}
\hline
Both&\#LD&\#$f_3$&Only LD&\#&Only $f_3$&\#&Neither\\
\hline
Adygei&205&139&BiakaPygmy&81&French&99&Basque\\
Balochi&123&204&Colombian&5&Han&13&Dai\\
BantuKenya&30&182&Druze&128&Italian&46&Hezhen\\
BantuSouthAfrica&27&11&Japanese&1&Orcadian&1&Karitiana\\
Bedouin&300&63&Kalash&20&Tujia&8&Lahu\\
Brahui&363&16&MbutiPygmy&77&Tuscan&59&Mandenka\\
Burusho&450&377&Melanesian&96&&&Miao\\
Cambodian&266&158&Pima&489&&&Naxi\\
Daur&29&8&San&155&&&Papuan\\
Han-NChina&1&77&Sardinian&45&&&She\\
Hazara&699&593&Yakut&435&&&Surui\\
Makrani&173&163&&&&&Yi\\
Maya&784&124&&&&&Yoruba\\
Mongola&76&385&&&&&\\
Mozabite&313&107&&&&&\\
Oroqen&68&5&&&&&\\
Palestinian&308&64&&&&&\\
Pathan&113&348&&&&&\\
Russian&158&153&&&&&\\
Sindhi&264&366&&&&&\\
Tu&22&315&&&&&\\
Uygur&428&616&&&&&\\
Xibo&101&335&&&&&\\
\hline
\end{tabular}
}
\begin{flushleft}
We ran both \ALDER and the 3-population test for admixture on each of
the 53 HGDP populations using all pairs of other populations as
references.  We group the populations according to whether or not each
test methodology produced at least one test identifying them as admixed;
for each population, we list the number of reference pairs with which with each method (abbreviated ``LD'' and
``$f_3$'') detected admixture.  We used a significance threshold of $p < 0.05$ after
multiple-hypothesis correction.
\end{flushleft}
\label{tab:HGDP_LD_vs_f3}
\end{table}

\newpage
\begin{table}[H]
\caption{\bf{Amplitudes and dates from weighted LD curves for Sardinian using various reference pairs.\newline}}
{\small\sffamily
\begin{tabular}{llll}
\hline
Ref 1 & Ref 2 & Weighted LD amplitude & Date estimate \\
\hline
CEU & YRI & 0.00003192 $\pm$ 0.00000903 & 48 $\pm$ 10 \\
CHB & YRI & 0.00001738 $\pm$ 0.00000679 & 34 $\pm$ 8 \\
CEU & CHB & 0.00000873 $\pm$ 0.00000454 & 52 $\pm$ 21 \\
\hline
\end{tabular}
}
\begin{flushleft}
Data are shown from \ALDER fits to weighted LD curves computed using
Sardinian as the test population and pairs of HapMap CEU, YRI, and CHB as
the references.  Date estimates are in generations.  We omitted chromosome 8 from the analysis because of
anomalous long-range LD.  Curves $\hat{a}(d)$ were fit for $d > 1.2$ cM, the
extent of LD correlation between Sardinian and CEU computed by \ALDER.
\end{flushleft}
\label{tab:Sardinian_amp_tots}
\end{table}

\newpage
\section*{Supporting Information}

\setcounter{figure}{0}
\renewcommand{\thefigure}{S\arabic{figure}}

\begin{figure}[H]
\begin{center}
\includegraphics[width=\textwidth]{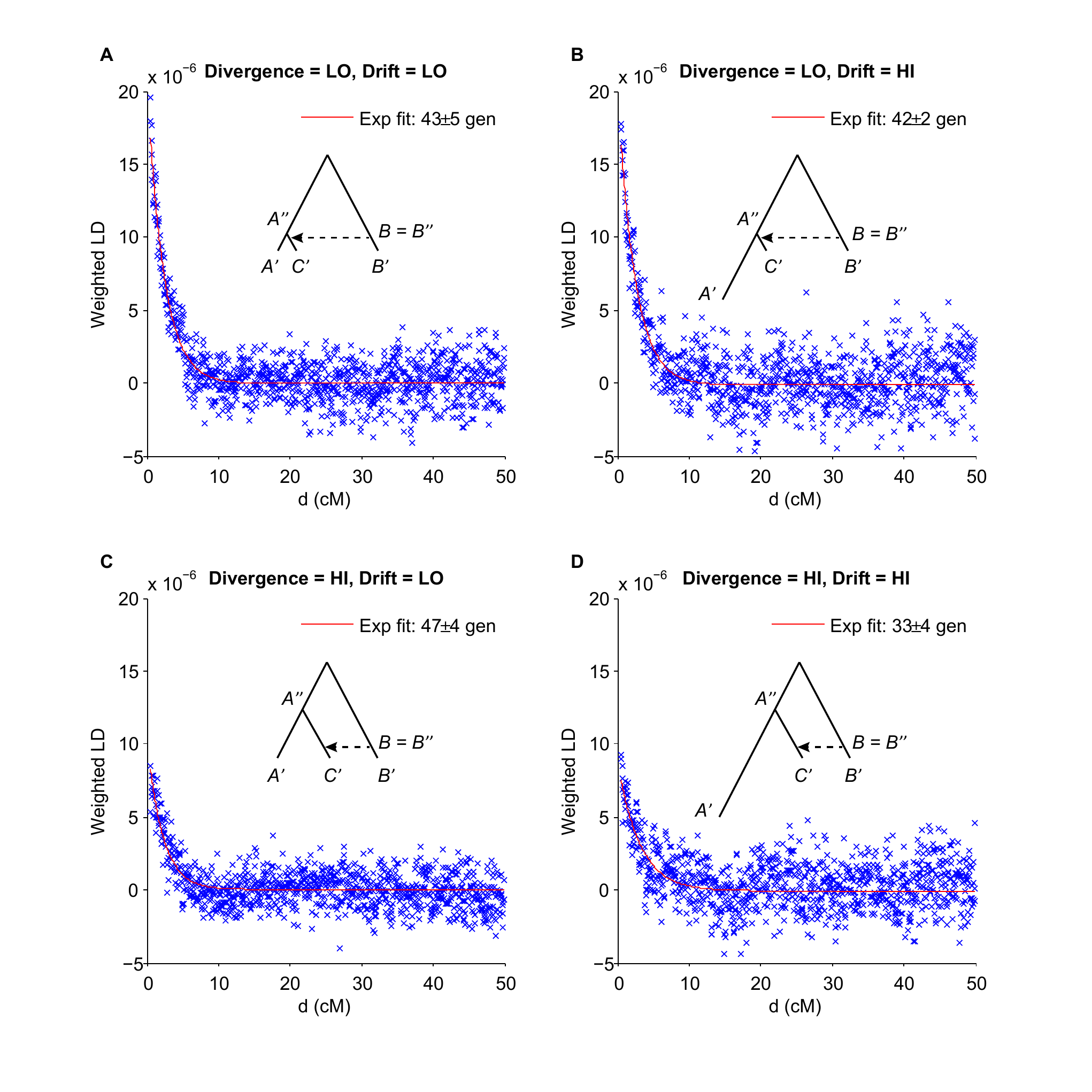}
\end{center}
\caption{Weighted LD curves from four coalescent simulations of
  admixture scenarios with varying divergence times and drift between
  the reference population $A'$ and the true mixing population.  In
  each case, gene flow occurred 40 generations ago.  In the low-divergence scenarios, the split point $A''$ is immediately prior to gene flow, while in the high-divergence scenarios, $A''$ is halfway up the tree (520 generations ago).  The high-drift scenarios are distinguished from the low-drift scenarios by a 20-fold reduction in population size for the past 40 generations.  Standard errors
  shown are \ALDER's jackknife estimates of its own error on a single
  simulation.}
\label{fig:macs_sims:amp_diverged_refs}
\end{figure}

\newpage
\begin{figure}[H]
\begin{center}
\includegraphics[scale=1.5]{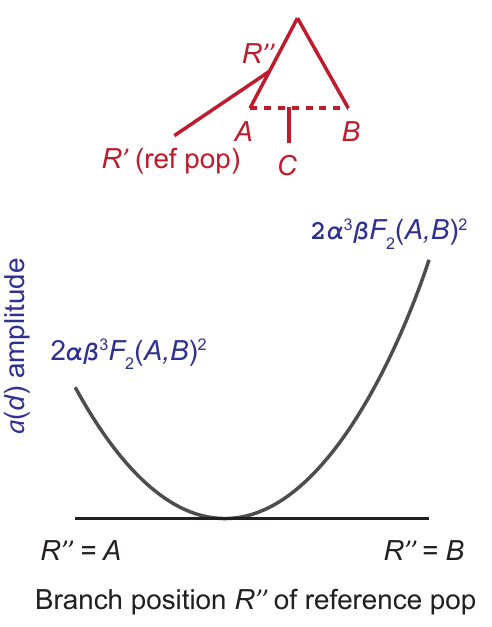}
\end{center}
\caption{Dependence of single-reference weighted LD amplitude on the
  reference population.  When taking weights as allele frequency
  differences between the admixed population and a single reference
  population $R'$, the weighted LD curve $a(d)$ has expected amplitude
  proportional to $(\alpha F_2(A,R'') - \beta F_2(B,R''))^2$, where
  $R''$ is the point along the $A$--$B$ lineage at which the reference
  population branches.  Note in particular that as $R''$ varies from
  $A$ to $B$, the amplitude traces out a parabola that starts at
  $2\alpha\beta^3 F_2(A,B)^2$, decreases to a minimum value of 0, and
  increases to $2\alpha^3\beta F_2(A,B)^2$.}
\label{fig:alder_parabola}
\end{figure}

\newpage
\begin{figure}[H]
\begin{center}
\includegraphics[scale=0.75]{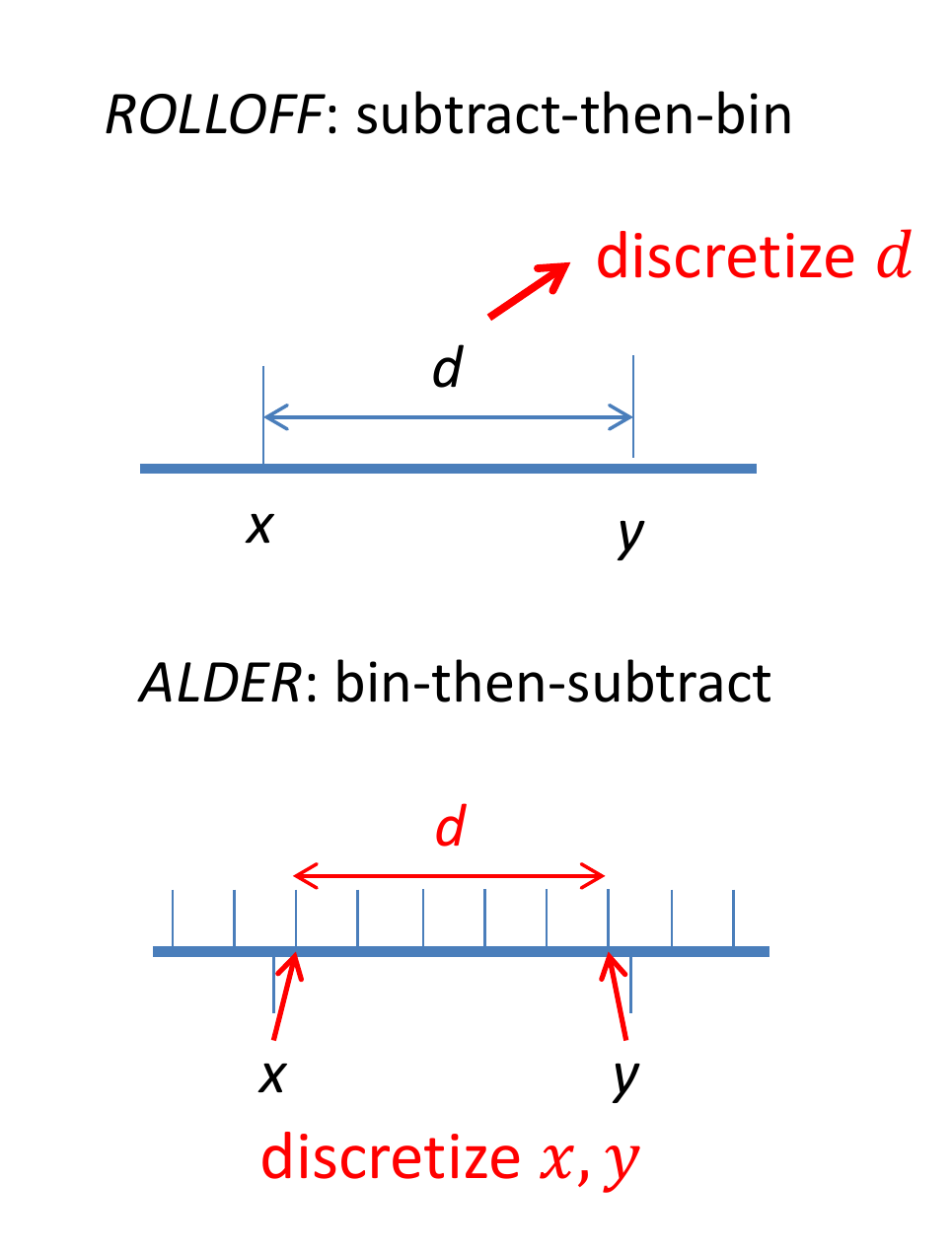}
\end{center}
\caption{Comparison of binning procedures used by \ROLLOFF and \ALDER.  Instead of discretizing inter-SNP distances, \ALDER discretizes the genetic map before subtracting SNP coordinates.}
\label{fig:binning}
\end{figure}

\newpage
\begin{figure}[H]
\begin{center}
\includegraphics[width=\textwidth]{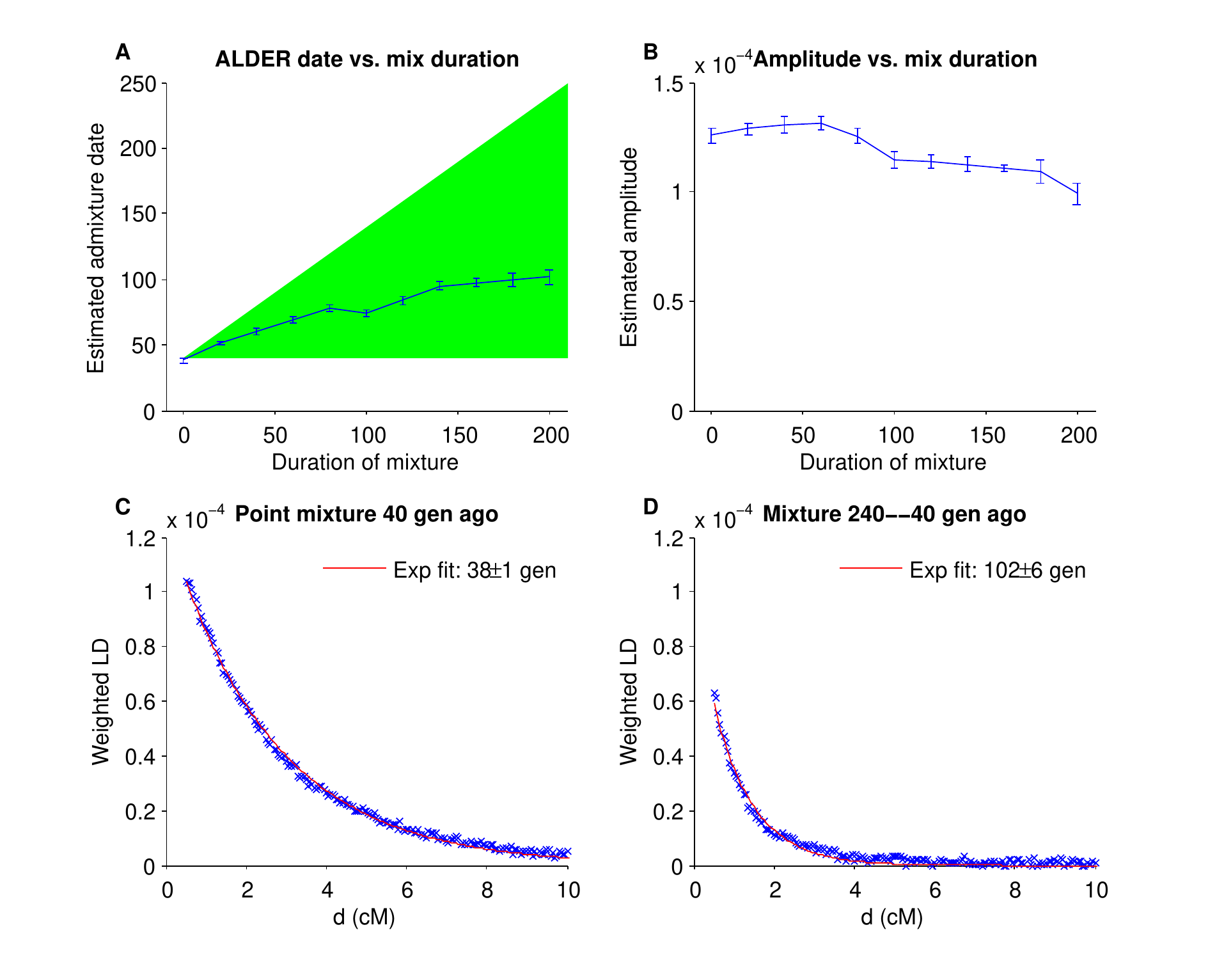}
\end{center}
\caption{Weighted LD curve parameters from coalescent simulations of
  continuous admixture.  In each simulation the mixed population
  receives $40\%$ of its ancestry through continuous gene flow over a
  period of 0--200 generations ending 40 generations ago.  Panels (A)
  and (B) show the admixture dates and weighted LD amplitudes computed
  by \ALDER for each of 11 simulations (varying the duration of
  mixture from 0 to 200 in increments of 20).  Panels (C) and (D) show
  the curves and exponential fits for mixture durations at the two
  extremes.  Standard errors shown are \ALDER's jackknife estimates of
  its own error on a single simulation.}
\label{fig:macs_sims:continuous}
\end{figure}

\newpage
\begin{figure}[H]
\begin{center}
\includegraphics[width=\textwidth]{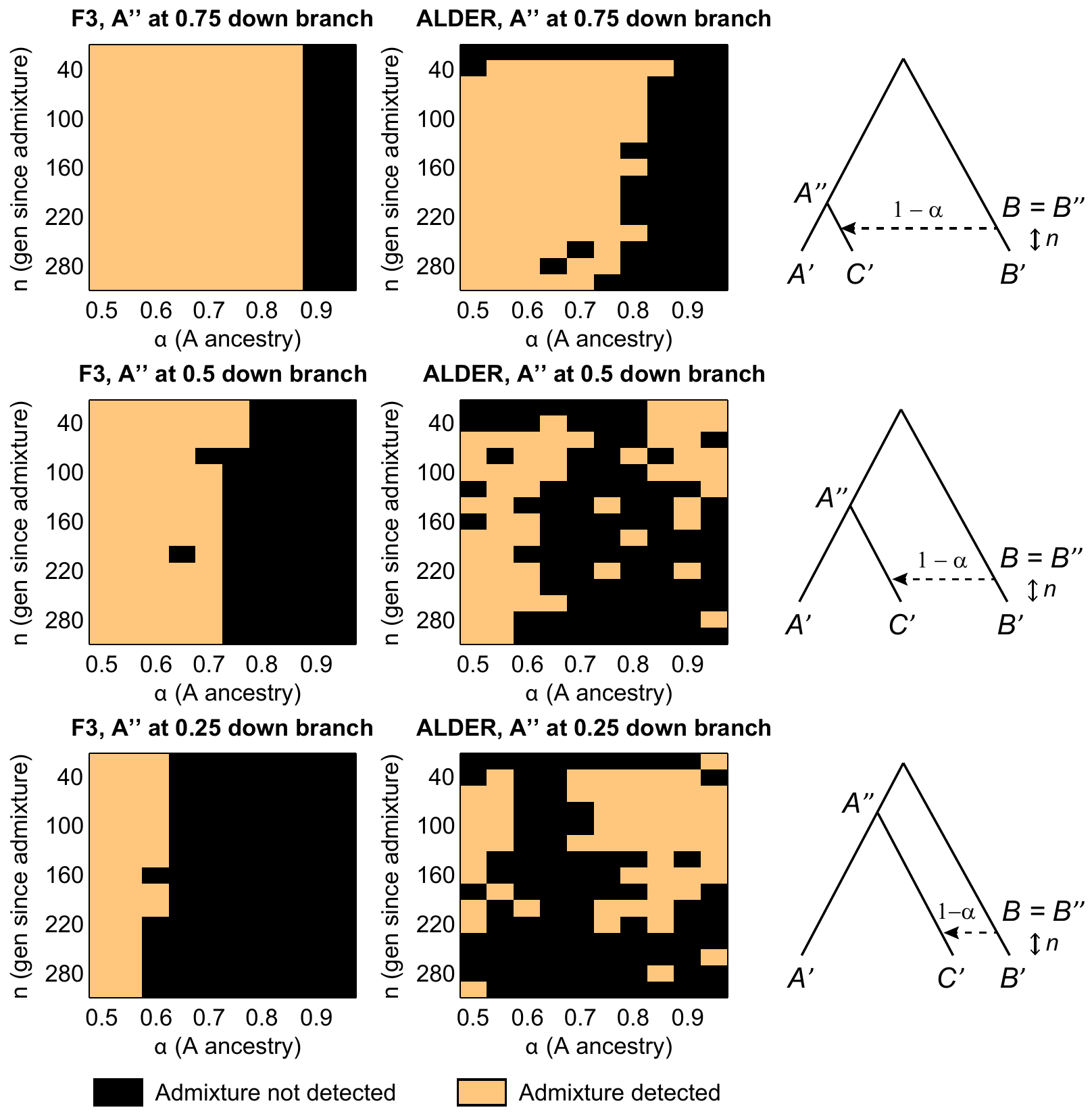}
\end{center}
\caption{Coalescent simulations comparing the sensitivities of the
  3-population moment-based test for admixture ($f_3$) and the
  LD-based test implemented in \ALDER.  We varied three parameters:
  the age of the branch point $A''$, the date $n$ of gene flow, and
  the fraction $\alpha$ of $A$ ancestry.}
\label{fig:macs_sims:f3_vs_alder}
\end{figure}

\newpage
\begin{figure}[H]
\begin{center}
\includegraphics[width=\textwidth]{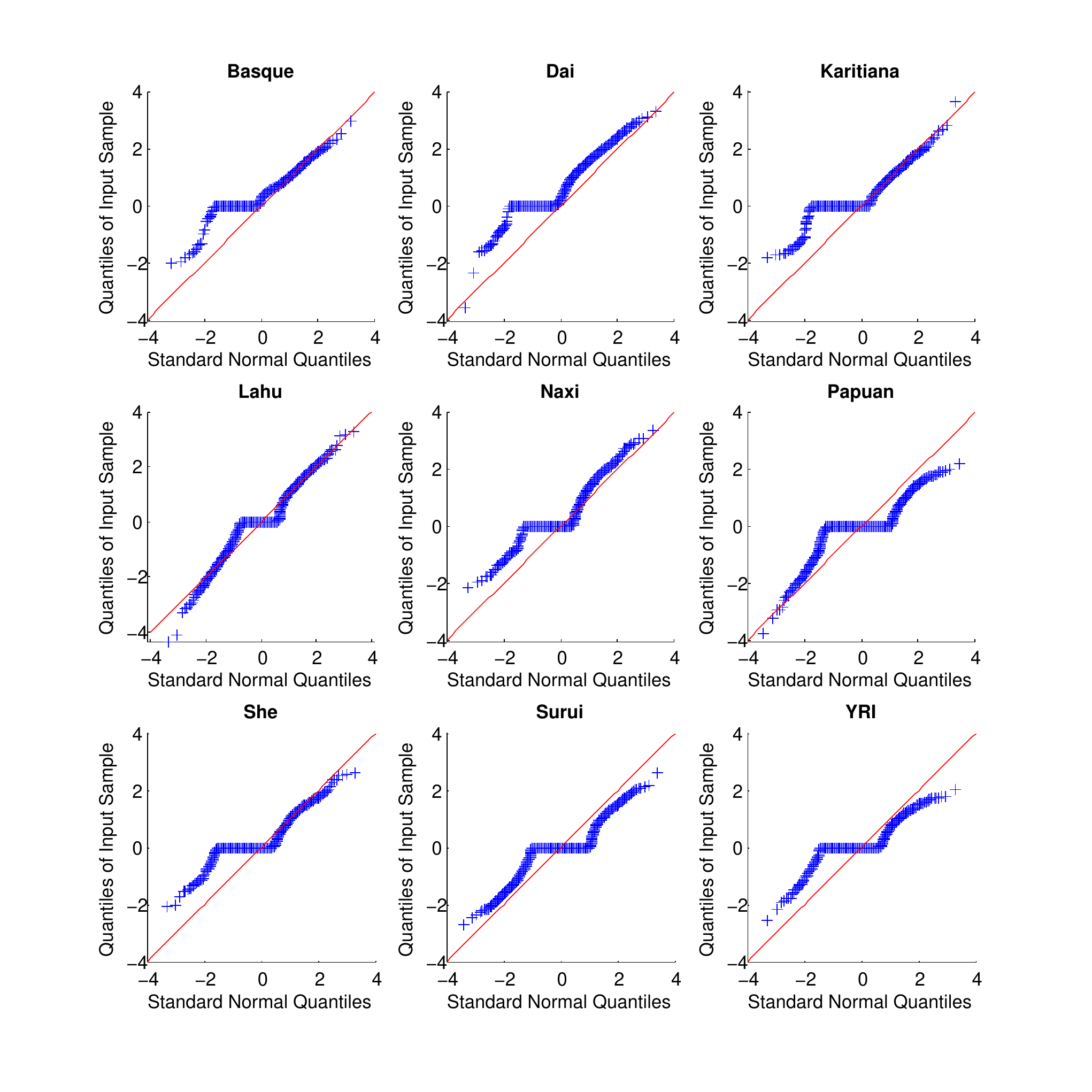}
\end{center}
\caption{Q-Q plots comparing \ALDER $z$-scores to standard normal on
  null examples.  We show results from nine HGDP populations that
  neither \ALDER nor the 3-population test found to be admixed.  We
  are interested in values of $z > 0$; the Q-Q plots show that these
  values follow the standard normal reasonably well, tending to err on the
  conservative side.
}
\label{fig:QQ_plots}
\end{figure}

\newpage
\begin{figure}[H]
\begin{center}
\includegraphics[scale=1.5]{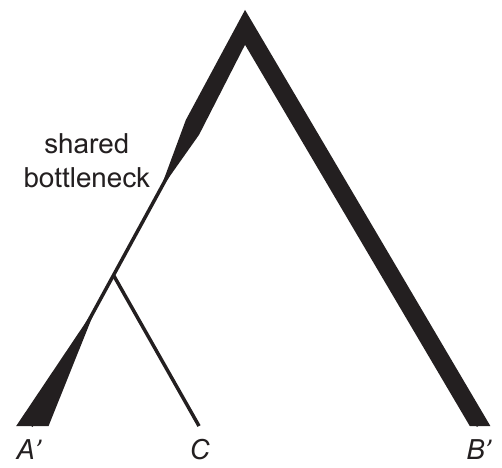}
\end{center}
\caption{Non-admixture-related demography producing weighted LD
  curves. The test population is $C$ and references are $A'$ and $B'$;
  the common ancestor of $A'$ and $C$ experienced a recent bottleneck
  from which $C$ has not yet recovered, leaving long-range LD in $C$
  that is potentially correlated to all three possible weighting
  schemes ($A'$--$B'$, $A'$--$C$, and $B'$--$C$).}
\label{fig:pathological_demography}
\end{figure}

\setcounter{table}{0}
\renewcommand{\thetable}{S\arabic{table}}

\newpage
\begin{table}[H]
\caption{Dates of admixture estimated for simulated 75\% YRI / 25\% CEU mixtures.\newline}
{\small\sffamily
\begin{tabular}{llrrrrr}
\hline
Ref 1&Ref 2&10&20&50&100&200\\
\hline
Yoruba&French&9$\pm$1&20$\pm$1&49$\pm$2&107$\pm$5&195$\pm$9\\
Yoruba&Han&9$\pm$1&21$\pm$1&50$\pm$2&107$\pm$6&191$\pm$12\\
Yoruba&Papuan&9$\pm$1&21$\pm$1&49$\pm$3&118$\pm$8&223$\pm$23\\
San&French&9$\pm$1&20$\pm$1&50$\pm$2&109$\pm$4&197$\pm$15\\
San&Han&9$\pm$0&21$\pm$1&51$\pm$3&111$\pm$4&194$\pm$16\\
San&Papuan&9$\pm$1&21$\pm$1&51$\pm$3&115$\pm$6&209$\pm$16\\
Yoruba&&9$\pm$1&21$\pm$1&48$\pm$2&107$\pm$5&181$\pm$17\\
San&&9$\pm$1&20$\pm$2&56$\pm$7&139$\pm$22&213$\pm$97\\
French&&9$\pm$1&20$\pm$1&50$\pm$2&108$\pm$3&194$\pm$9\\
Han&&9$\pm$0&21$\pm$1&52$\pm$2&110$\pm$6&192$\pm$17\\
Papuan&&9$\pm$1&21$\pm$1&53$\pm$3&125$\pm$8&217$\pm$26\\
\hline
\end{tabular}
}
\begin{flushleft}
We simulated scenarios in which admixture occurred 10, 20, 50, 100, or
200 generations ago and show results from runs of \ALDER using various
references.  Rows in which only one reference is listed indicate runs
using the admixed population itself as one reference.
Note that standard errors shown are \ALDER's jackknife estimates of
its own error on a single simulation (not standard errors from
averaging over multiple simulations).
\end{flushleft}
\label{tab:sim_dates_75_25}
\end{table}

\newpage
\begin{table}[H]
\caption{Dates of admixture estimated for simulated 90\% YRI / 10\% CEU mixtures.\newline}
{\small\sffamily
\begin{tabular}{llrrrrr}
\hline
Ref 1&Ref 2&10&20&50&100&200\\
\hline
Yoruba&French&10$\pm$0&21$\pm$1&50$\pm$2&107$\pm$7&193$\pm$19\\
Yoruba&Han&10$\pm$0&20$\pm$1&51$\pm$2&109$\pm$10&220$\pm$32\\
Yoruba&Papuan&10$\pm$0&22$\pm$1&53$\pm$3&111$\pm$11&233$\pm$65\\
San&French&10$\pm$0&21$\pm$1&51$\pm$2&112$\pm$6&223$\pm$19\\
San&Han&10$\pm$0&21$\pm$1&52$\pm$3&121$\pm$5&254$\pm$40\\
San&Papuan&11$\pm$0&23$\pm$1&53$\pm$3&126$\pm$8&287$\pm$56\\
Yoruba&&9$\pm$1&20$\pm$2&55$\pm$7&100$\pm$27&363$\pm$183\\
San&&98$\pm$87&56$\pm$28&94$\pm$69&2$\pm$0&9$\pm$5\\
French&&10$\pm$0&21$\pm$1&51$\pm$2&107$\pm$5&217$\pm$13\\
Han&&11$\pm$0&21$\pm$1&52$\pm$2&111$\pm$7&234$\pm$25\\
Papuan&&11$\pm$0&23$\pm$1&56$\pm$3&117$\pm$8&256$\pm$47\\
\hline
\end{tabular}
}
\begin{flushleft}
We simulated scenarios in which admixture occurred 10, 20, 50, 100, or
200 generations ago and show results from runs of \ALDER using various
references.  Rows in which only one reference is listed indicate runs
using the admixed population itself as one reference.
Note that standard errors shown are \ALDER's jackknife estimates of
its own error on a single simulation (not standard errors from
averaging over multiple simulations).
\end{flushleft}
\label{tab:sim_dates_90_10}
\end{table}

\newpage
\begin{table}[H]
\caption{Amplitudes of weighted LD curves (multiplied by $10^6$) for simulated 75\% YRI / 25\% CEU mixtures.\newline}
{\small\sffamily
\begin{tabular}{llrrrrrr}
\hline
Ref 1&Ref 2&Expected&10 gen&20 gen&50 gen&100 gen&200 gen\\
\hline
Yoruba&French&1173&1139$\pm$20&1203$\pm$40&1188$\pm$54&1283$\pm$100&1202$\pm$88\\
Yoruba&Han&693&678$\pm$17&717$\pm$28&711$\pm$43&774$\pm$73&716$\pm$74\\
Yoruba&Papuan&602&598$\pm$13&631$\pm$23&595$\pm$34&775$\pm$96&835$\pm$152\\
San&French&1017&981$\pm$23&1028$\pm$34&1044$\pm$49&1128$\pm$70&1037$\pm$130\\
San&Han&574&556$\pm$18&590$\pm$24&604$\pm$42&667$\pm$39&626$\pm$65\\
San&Papuan&491&487$\pm$17&514$\pm$20&503$\pm$34&589$\pm$45&574$\pm$60\\
Yoruba&&75&77$\pm$2&81$\pm$4&74$\pm$4&83$\pm$6&71$\pm$13\\
San&&40&40$\pm$3&42$\pm$3&50$\pm$6&66$\pm$13&43$\pm$34\\
French&&655&626$\pm$12&660$\pm$21&666$\pm$31&721$\pm$42&656$\pm$49\\
Han&&312&304$\pm$10&324$\pm$14&332$\pm$23&364$\pm$25&332$\pm$36\\
Papuan&&252&256$\pm$9&273$\pm$13&267$\pm$17&331$\pm$34&314$\pm$55\\
\hline
\end{tabular}
}
\begin{flushleft}
We simulated scenarios in which admixture occurred 10, 20, 50, 100, or
200 generations ago and show results from runs of \ALDER using various
references.  Rows in which only one reference is listed indicate runs
using the admixed population itself as one reference.  Expected
amplitudes were computed according to formulas \eqref{eq:amp_tworef}
and \eqref{eq:amp_oneref}.
Note that standard errors shown are \ALDER's jackknife estimates of
its own error on a single simulation (not standard errors from
averaging over multiple simulations).
\end{flushleft}
\label{tab:sim_amplitudes_75_25}
\end{table}

\newpage
\begin{table}[H]
\caption{Amplitudes of weighted LD curves (multiplied by $10^6$) for simulated 90\% YRI / 10\% CEU mixtures.\newline}
{\small\sffamily
\begin{tabular}{llrrrrrr}
\hline
Ref 1&Ref 2&Expected&10 gen&20 gen&50 gen&100 gen&200 gen\\
\hline
Yoruba&French&563&587$\pm$27&579$\pm$26&550$\pm$25&600$\pm$43&562$\pm$96\\
Yoruba&Han&333&353$\pm$20&336$\pm$15&339$\pm$17&381$\pm$49&456$\pm$128\\
Yoruba&Papuan&289&307$\pm$19&303$\pm$16&309$\pm$18&343$\pm$54&426$\pm$248\\
San&French&488&522$\pm$25&512$\pm$22&488$\pm$25&519$\pm$28&625$\pm$89\\
San&Han&276&305$\pm$18&291$\pm$12&289$\pm$16&338$\pm$23&464$\pm$132\\
San&Papuan&236&266$\pm$18&262$\pm$13&254$\pm$12&306$\pm$38&486$\pm$186\\
Yoruba&&6&6$\pm$1&6$\pm$1&7$\pm$1&7$\pm$3&44$\pm$89\\
San&&1&16$\pm$15&8$\pm$3&10$\pm$7&-0$\pm$0&-1$\pm$1\\
French&&454&473$\pm$19&471$\pm$18&450$\pm$19&481$\pm$19&566$\pm$55\\
Han&&250&268$\pm$13&261$\pm$10&264$\pm$11&288$\pm$23&369$\pm$68\\
Papuan&&212&231$\pm$14&233$\pm$13&243$\pm$11&276$\pm$35&366$\pm$125\\
\hline
\end{tabular}
}
\begin{flushleft}
We simulated scenarios in which admixture occurred 10, 20, 50, 100, or
200 generations ago and show results from runs of \ALDER using various
references.  Rows in which only one reference is listed indicate runs
using the admixed population itself as one reference.  Expected
amplitudes were computed according to formulas \eqref{eq:amp_tworef}
and \eqref{eq:amp_oneref}.
Note that standard errors shown are \ALDER's jackknife estimates of
its own error on a single simulation (not standard errors from
averaging over multiple simulations).
\end{flushleft}
\label{tab:sim_amplitudes_90_10}
\end{table}

\newpage
\begin{table}[H]
\caption{Mixture fraction lower bounds on simulated 75\% YRI / 25\% CEU mixtures.\newline}
{\small\sffamily
\begin{tabular}{lrrrrr}
\hline
Ref&10&20&50&100&200\\
\hline
French&24.6$\pm$0.3&25.7$\pm$0.5&25.7$\pm$0.7&27.0$\pm$1.0&25.2$\pm$1.3\\
Russian&23.8$\pm$0.3&24.9$\pm$0.5&24.8$\pm$0.7&25.6$\pm$0.8&25.3$\pm$1.0\\
Sardinian&21.3$\pm$0.3&21.9$\pm$0.5&22.0$\pm$0.6&23.6$\pm$0.9&22.3$\pm$1.1\\
Kalash&14.7$\pm$0.2&15.5$\pm$0.4&15.5$\pm$0.5&16.4$\pm$0.6&15.6$\pm$0.9\\
Yoruba&73.6$\pm$0.7&74.8$\pm$0.4&74.0$\pm$0.6&76.2$\pm$1.3&73.8$\pm$3.4\\
Mandenka&50.5$\pm$0.6&51.2$\pm$1.0&50.4$\pm$1.4&54.9$\pm$2.0&60.8$\pm$5.6\\
\hline
\end{tabular}
}
\begin{flushleft}
We simulated scenarios in which admixture occurred 10, 20, 50, 100, or
200 generations ago and show results from runs of \ALDER using various
single references.  The first four rows are European surrogates and
give lower bounds on the amount of CEU ancestry (25\%); the last two
are African surrogates and give lower bounds on the amount of YRI
ancestry (75\%).
Note that standard errors shown are \ALDER's jackknife estimates of
its own error on a single simulation (not standard errors from
averaging over multiple simulations).
\end{flushleft}
\label{tab:sim_mix_props_75_25}
\end{table}

\newpage
\begin{table}[H]
\caption{Mixture fraction lower bounds on simulated 90\% YRI / 10\% CEU mixtures.\newline}
{\small\sffamily
\begin{tabular}{lrrrrr}
\hline
Ref&10&20&50&100&200\\
\hline
French&10.5$\pm$0.4&10.5$\pm$0.3&9.9$\pm$0.3&10.6$\pm$0.4&12.3$\pm$1.0\\
Russian&10.2$\pm$0.3&10.0$\pm$0.3&9.7$\pm$0.3&10.3$\pm$0.5&11.8$\pm$0.9\\
Sardinian&9.3$\pm$0.3&9.2$\pm$0.3&8.7$\pm$0.3&9.5$\pm$0.4&10.3$\pm$1.2\\
Kalash&7.2$\pm$0.3&7.0$\pm$0.3&6.8$\pm$0.2&7.4$\pm$0.4&8.9$\pm$0.8\\
Yoruba&89.1$\pm$1.0&89.1$\pm$1.1&90.1$\pm$1.5&89.4$\pm$3.7&98.5$\pm$2.0\\
Mandenka&18.2$\pm$2.3&17.3$\pm$2.5&19.5$\pm$4.8&63.1$\pm$25.5&30.7$\pm$220.4\\
\hline
\end{tabular}
}
\begin{flushleft}
We simulated scenarios in which admixture occurred 10, 20, 50, 100, or
200 generations ago and show results from runs of \ALDER using various
single references.  The first four rows are European surrogates and
give lower bounds on the amount of CEU ancestry (10\%); the last two
are African surrogates and give lower bounds on the amount of YRI
ancestry (90\%).
Note that standard errors shown are \ALDER's jackknife estimates of
its own error on a single simulation (not standard errors from
averaging over multiple simulations).
\end{flushleft}
\label{tab:sim_mix_props_90_10}
\end{table}

\newpage
\begin{table}[H]
\caption{Dates of admixture estimated for simulated 75\% YRI / 25\% CEU mixtures.\newline}
{\small\sffamily
\begin{tabular}{lrrrrr}
\hline
\multicolumn{6}{l}{Yoruba--French references}\\
\hline
Samples&10 gen&20 gen&50 gen&100 gen&200 gen\\
5&12$\pm$2&18$\pm$2&55$\pm$3&103$\pm$7&258$\pm$24\\
10&10$\pm$1&19$\pm$2&50$\pm$2&105$\pm$7&236$\pm$24\\
20&10$\pm$1&20$\pm$1&52$\pm$2&104$\pm$5&223$\pm$16\\
50&9$\pm$0&20$\pm$1&52$\pm$1&96$\pm$2&186$\pm$10\\
100&10$\pm$0&20$\pm$0&52$\pm$1&101$\pm$2&210$\pm$9\\
\hline
\multicolumn{6}{l}{San--Han references}\\
\hline
Samples&10 gen&20 gen&50 gen&100 gen&200 gen\\
5&12$\pm$2&18$\pm$2&58$\pm$5&107$\pm$11&283$\pm$73\\
10&10$\pm$1&19$\pm$2&54$\pm$3&114$\pm$8&219$\pm$64\\
20&10$\pm$1&21$\pm$1&55$\pm$2&115$\pm$6&219$\pm$46\\
50&9$\pm$0&21$\pm$1&54$\pm$1&107$\pm$5&213$\pm$20\\
100&9$\pm$0&21$\pm$1&53$\pm$1&105$\pm$5&216$\pm$13\\
\hline
\end{tabular}
}
\begin{flushleft}
We simulated scenarios in which admixture occurred 10, 20, 50, 100, or
200 generations ago and show results from runs of \ALDER using varying
numbers of admixed samples.  Note that standard errors shown are
\ALDER's jackknife estimates of its own error on a single simulation
(not standard errors from averaging over multiple simulations).
\end{flushleft}
\label{tab:sim_vary_nadmix_dates_75_25}
\end{table}

\newpage
\begin{table}[H]
\caption{Dates of admixture estimated for simulated 90\% YRI / 10\% CEU mixtures.\newline}
{\small\sffamily
\begin{tabular}{lrrrrr}
\hline
\multicolumn{6}{l}{Yoruba--French references}\\
\hline
Samples&10 gen&20 gen&50 gen&100 gen&200 gen\\
5&11$\pm$2&21$\pm$2&52$\pm$6&101$\pm$17&253$\pm$42\\
10&11$\pm$1&19$\pm$1&48$\pm$4&94$\pm$8&241$\pm$46\\
20&11$\pm$1&21$\pm$1&48$\pm$3&102$\pm$8&209$\pm$30\\
50&11$\pm$0&21$\pm$1&48$\pm$2&98$\pm$5&202$\pm$21\\
100&10$\pm$0&20$\pm$1&50$\pm$1&99$\pm$4&185$\pm$15\\
\hline
\multicolumn{6}{l}{San--Han references}\\
\hline
Samples&10 gen&20 gen&50 gen&100 gen&200 gen\\
5&14$\pm$2&22$\pm$3&63$\pm$8&110$\pm$30&335$\pm$91\\
10&12$\pm$1&20$\pm$2&54$\pm$4&110$\pm$15&265$\pm$55\\
20&12$\pm$1&21$\pm$1&52$\pm$4&131$\pm$15&234$\pm$33\\
50&11$\pm$0&20$\pm$1&53$\pm$4&122$\pm$8&221$\pm$23\\
100&11$\pm$0&20$\pm$0&53$\pm$3&109$\pm$5&219$\pm$10\\
\hline
\end{tabular}
}
\begin{flushleft}
We simulated scenarios in which admixture occurred 10, 20, 50, 100, or
200 generations ago and show results from runs of \ALDER using varying
numbers of admixed samples.  Note that standard errors shown are
\ALDER's jackknife estimates of its own error on a single simulation
(not standard errors from averaging over multiple simulations).
\end{flushleft}
\label{tab:sim_vary_nadmix_dates_90_10}
\end{table}

\newpage
\begin{table}[H]
\caption{Effect of SNP ascertainment on date estimates.\newline}
{\small\sffamily
\begin{tabular}{lllrrrr}
\hline
Mixed pop&Ref 1&Ref 2&French asc&Han asc&San asc&Yoruba asc\\
\hline
Burusho&French&Han&47$\pm$12&51$\pm$13&56$\pm$10&41$\pm$10\\
Uygur&French&Han&15$\pm$2&14$\pm$2&13$\pm$2&16$\pm$2\\
Hazara&French&Han&22$\pm$2&22$\pm$3&23$\pm$2&22$\pm$3\\
Melanesian&Dai&Papuan&93$\pm$24&62$\pm$15&76$\pm$13&70$\pm$18\\
Bedouin&French&Yoruba&27$\pm$3&23$\pm$3&23$\pm$3&24$\pm$3\\
MbutiPygmy&San&Yoruba&33$\pm$12&33$\pm$6&41$\pm$14&30$\pm$8\\
BiakaPygmy&San&Yoruba&39$\pm$6&50$\pm$14&35$\pm$6&36$\pm$7\\
\hline
\end{tabular}
}
\begin{flushleft}
We compared dates of admixture estimated by \ALDER on a variety of
test triples from the HGDP using SNPs ascertained as heterozygous in full genome sequences of one French, Han, San, and Yoruba
individual (Panels 1, 2, 4, and 5 of the Affymetrix Human
Origins Array \cite{draft7}).  Standard errors are from a jackknife
over the 22 autosomes.
\end{flushleft}
\label{tab:ascertainment_effect_on_date}
\end{table}

\newpage
\begin{table}[H]
\caption{Effect of SNP ascertainment on weighted LD curve amplitudes (multiplied by $10^6$).\newline}
{\small\sffamily
\begin{tabular}{lllrrrr}
\hline
Mixed pop&Ref 1&Ref 2&French asc&Han asc&San asc&Yoruba asc\\
\hline
Burusho&French&Han&180$\pm$44&171$\pm$53&61$\pm$11&65$\pm$15\\
Uygur&French&Han&360$\pm$28&304$\pm$29&102$\pm$7&161$\pm$19\\
Hazara&French&Han&442$\pm$31&436$\pm$48&146$\pm$10&203$\pm$21\\
Melanesian&Dai&Papuan&868$\pm$277&559$\pm$150&207$\pm$51&312$\pm$91\\
Bedouin&French&Yoruba&227$\pm$32&196$\pm$25&104$\pm$11&146$\pm$13\\
MbutiPygmy&San&Yoruba&64$\pm$23&78$\pm$14&83$\pm$26&82$\pm$18\\
BiakaPygmy&San&Yoruba&104$\pm$19&133$\pm$46&90$\pm$15&103$\pm$22\\
\hline
\end{tabular}
}
\begin{flushleft}
We compared amplitudes of weighted LD curves fitted on a variety of
test triples from the HGDP using SNPs ascertained as heterozygous in full genome sequences of one French, Han, San, and Yoruba
individual (Panels 1, 2, 4, and 5 of the Affymetrix Human
Origins Array \cite{draft7}).  Standard errors are from a jackknife
over the 22 autosomes.
\end{flushleft}
\label{tab:ascertainment_effect_on_amp}
\end{table}

\setcounter{figure}{0}
\renewcommand{\thefigure}{S\arabic{figure}}
\renewcommand{\figurename}{File}

\newpage
\begin{figure}[H]
\caption{{\bf Unbiased polyache estimator for weighted LD using the admixed population itself as one reference.}}
\begin{center}
\includegraphics[width=\textwidth]{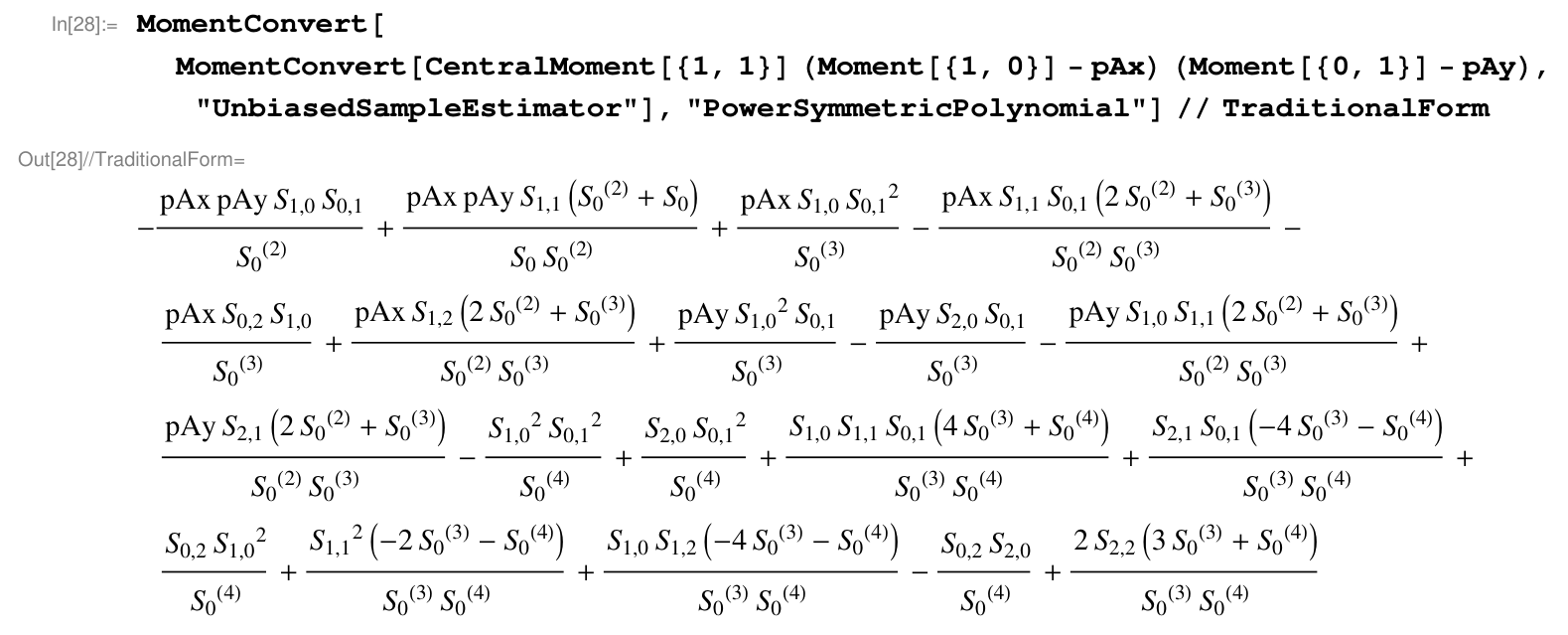}
\end{center}
\label{fig:single_anc_unbiased_estimator}
\end{figure}
\noindent
Mathematica code and output are shown for computing the
polyache statistic that estimates the one-reference weighted LD,
$E[(X-\mu_x)(Y-\mu_y)(\mu_x - p_A(x))(\mu_y - p_A(y))]$, where
$p_A(\cdot)$ are allele frequencies of the single reference population
and $\mu_x$ and $\mu_y$ denote allele frequencies of the admixed
population.  In the above, $S_0^{(k)} := m(m-1)\cdots(m-k+1)$ and
$S_{r,s} := \sum_{i=1}^m X_i^rY_i^s$, where $m$ is the number of
admixed samples and $i$ ranges over the admixed individuals, which
have allele counts $X_i$ and $Y_i$ at sites $x$ and $y$.

\newpage
\begin{figure}[H]
\caption{{\bf FFT computation of weighted LD.}}
\label{file:fast_weighted_ld}
\end{figure}

\setcounter{equation}{0}
\def\sumapproxd{\sum_{|x-y| \approx d}}

In this note we describe how to compute weighted LD (aggregated over
distance bins) in time
\[
O(m(S+B\log B)),
\]
where $m$ is the number of admixed individuals, $S$ is the number of
SNPs, and $B$ is the number of bins needed to span the chromosomes.
In contrast, the direct method of computing pairwise LD for each
individual SNP pair requires $O(mS^2)$ time.  In practice our approach
offers speedups of over 1000x on typical data sets.  We further
describe a similar algorithm for computing the single-reference
weighted LD polyache statistic that runs in time
\[
O(m^2(S+B\log B))
\]
with the slight trade-off of ignoring SNPs with missing data.

Our method consists of three key steps: (1) split and factorize the
weighted LD product; (2) group factored terms by bin; and (3) apply
fast Fourier transform (FFT) convolution.  As a special case of this
approach, the first two ideas alone allow us to efficiently compute
the affine term (i.e., horizontal asymptote) of the weighted LD curve
using inter-chromosome SNP pairs.

\section*{Two-reference weighted LD}

We first establish notation.  Say we have an $S \times m$ genotype
array $\{c_{x,i}\}$ from an admixed population.  Assume for now that
there are no missing values, i.e.,
\[
c_{x,i} \in \{0, 1, 2\}
\]
for $x$ indexing SNPs by position on a genetic map and $i = 1, \dots,
m$ indexing individuals.  Given a set of weights $w_x$, one per SNP,
we wish to compute weighted LD of SNP pairs aggregated by inter-SNP
distance $d$:
\[
R(d) := \sum_{\substack{|x-y| \approx d\\x<y}} D_2(x,y)w_xw_y
= \frac{1}{2} \sumapproxd D_2(x,y)w_xw_y
\]
where $D_2$ is the sample covariance between genotypes at $x$ and $y$,
the diploid analog of the usual LD measure $D$:
\begin{eqnarray}
D_2(x,y) &:=& \frac{1}{m-1} \sum_{i=1}^m c_{x,i}c_{y,i} -
\frac{1}{m(m-1)} \sum_{i=1}^m c_{x,i} \sum_{j=1}^m c_{y,j} \nonumber \\
&=& \frac{1}{m-1} \sum_{i=1}^m c_{x,i}c_{y,i} - \frac{1}{m(m-1)} s_xs_y, \label{eq:diploidD}
\end{eqnarray}
where we have defined
\[
s_x := \sum_{i=1}^m c_{x,i}.
\]

Substituting for $D_2(x,y)$, we have
\begin{eqnarray}
R(d) &=& \frac{1}{2} \sumapproxd \left(
\frac{1}{m-1} \sum_{i=1}^m c_{x,i}c_{y,i} - \frac{1}{m(m-1)} s_xs_y
\right) w_xw_y \nonumber \\
&=& \left( \sum_{i=1}^m \frac{1}{2(m-1)} \sumapproxd c_{x,i}w_x \cdot c_{y,i}w_y \right)
- \frac{1}{2m(m-1)} \sumapproxd s_xw_x \cdot s_yw_y. \label{eq:m+1conv_nomissing}
\end{eqnarray}
We have thus rewritten $R(d)$ as a linear combination of $m+1$ terms
of the form
\[
\sumapproxd f(x)f(y).
\]
(The sum over $i$ consists of $m$ such terms, and the final term
accounts for one more.)

In general, sums of the form
\[
\sumapproxd f(x)g(y)
\]
can be efficiently computed by convolution if we first discretize the
genetic map on which the SNP positions $x$ and $y$ lie.  For
notational convenience, choose the distance scale such that a unit
distance corresponds to the desired bin resolution.  We will compute
\begin{equation} \label{eq:floor_convolution}
\sum_{\lfloor x \rfloor - \lfloor y \rfloor = d} f(x)g(y).
\end{equation}
That is, we divide the chromosome into bins of unit distance and
aggregate terms $f(x)g(y)$ by the distance between the bin centers of
$x$ and $y$.  Note that this procedure does not produce exactly the
same result as first subtracting the genetic positions and then
binning by $|x-y|$: with our approach, pairs $(x,y)$ that map to a
given bin can have actual distances that are off by as much as one
full bin width, versus half a bin width with the subtract-then-bin
approach.  However, we can compensate simply by doubling the bin
resolution.

To compute expression \eqref{eq:floor_convolution}, we write
\begin{eqnarray}
\sum_{\lfloor x \rfloor - \lfloor y \rfloor = d} f(x)g(y)
&=& \sum_{b=0}^B \sum_{\lfloor x \rfloor = b} \sum_{\lfloor y \rfloor = b-d} f(x)g(y) \nonumber \\
&=& \sum_{b=0}^B \left( \sum_{\lfloor x \rfloor = b} f(x) \right)
\left( \sum_{\lfloor y \rfloor = b-d} g(y) \right). \label{eq:agg_convolution}
\end{eqnarray}
Writing
\[
F(b) := \sum_{\lfloor x \rfloor = b} f(x), \quad G(b) := \sum_{\lfloor x \rfloor = b} g(x),
\]
expression \eqref{eq:agg_convolution} becomes
\[
\sum_{b=0}^B F(b) G(b-d) = (F \star G)(d),
\]
a cross-correlation of binned $f(x)$ and $g(y)$ terms.

Computationally, binning $f$ and $g$ to form $F$ and $G$ takes $O(S)$
time, after which the cross-correlation can be performed in $O(B\log B)$
time with a fast Fourier transform.  The full computation of the $m+1$
convolutions in equation \eqref{eq:m+1conv_nomissing} thus takes
$O(m(S + B\log B))$ time.  In practice we often have $B\log B < S$, in
which case the computation is linear in the data size $mS$.

One additional detail is that we usually want to compute the average
rather than the sum of the weighted LD contributions of the SNP pairs
in each bin; this requires normalizing by the number of pairs $(x,y)$
that map to each bin, which can be computed in an analogous manner
with one more convolution (setting $f \equiv 1$, $g \equiv 1$).
Finally, we note that our factorization and binning approach
immediately extends to computing weighted LD on inter-chromosome SNP
pairs (by putting all SNPs in a chromosome in the same bin), which
allows robust estimation of the horizontal asymptote of the weighted
LD curve.

\subsection*{Missing Data}

The calculations above assumed that the genotype array contained no
missing data, but in practice a fraction of the genotype values may be
missing.  The straightforward non-FFT computation has no difficulty
handling missing data, as each pairwise LD term $D_2(x,y)$ can be
calculated as a sample covariance over just the individuals
successfully genotyped at both $x$ and $y$.  Our algebraic
manipulation runs into trouble, however, because if $k$ individuals
have a missing value at either $x$ or $y$, then the sample covariance
contains denominators of the form $1/(m-k-1)$ and
$1/(m-k)(m-k-1)$---and $k$ varies depending on $x$ and $y$.

One way to get around this problem is simply to restrict the analysis
to sites with no missing values at the cost of slightly reduced power.
If a fraction $p$ of the SNPs contain at least one missing value, this
workaround reduces the number of SNP pairs available to $(1-p)^2$ of
the total, which is probably already acceptable in practice.

We can do better, however: in fact, with a little more algebra (but no
additional computational complexity), we can include all pairs of
sites $(x,y)$ for which at least one of the SNPs $x$, $y$ has no
missing values, bringing our coverage up to $1-p^2$.

We will need slightly more notation.  Adopting \texttt{eigenstrat}
format, we now let our genotype array consist of values
\[
c_{x,i} \in \{0, 1, 2, 9\}
\]
where 9 indicates a missing value.  (Thus, $\{c_{x,i}\}$ is exactly
the data that would be contained in a \texttt{.geno} file.)  For
convenience, we write
\[
c_{x,i}^{(0)} :=
\begin{cases}
  c_{x,i} & \mbox{if } c_{x,i} \in \{0, 1, 2\} \\
  0 & \mbox{otherwise.}
\end{cases}
\]
That is, $c_{x,i}^{(0)}$ replaces missing values with 0s.  As before
we set
\[
s_x := \sum_{i : c_{x,i} \ne 9} c_{x,i} = \sum_{i=1}^m c_{x,i}^{(0)}
\]
to be the sum of all non-missing values at $x$, which also equals the
sum of all $c_{x,i}^{(0)}$ because the missing values have been
0-replaced.  Finally, define
\[
k_x := \#\{i : c_{x,i} = 9\}
\]
to be the number of missing values at site $x$.

We now wish to compute aggregated weighted LD over pairs $(x,y)$ for
which at least one of $k_x$ and $k_y$ is 0.  Being careful not to
double-count, we have:
\begin{eqnarray}
R(d) &:=& \sum_{\substack{|x-y| \approx d\\x<y\\k_x=0 \text{ or } k_y=0}} D_2(x,y)w_xw_y \nonumber \\
&=& \frac{1}{2} \sum_{\substack{|x-y| \approx d\\k_x=0 \text{ and } k_y=0}} D_2(x,y)w_xw_y
+ \sum_{\substack{|x-y| \approx d\\k_x=0 \text{ and } k_y \ne 0}} D_2(x,y)w_xw_y \nonumber \\
&=& \sumapproxd \frac{I[k_x=0]}{1 + I[k_y=0]} D_2(x,y)w_xw_y, \label{eq:R_x_nomissing}
\end{eqnarray}
where the shorthand $I[\cdot]$ denotes a $\{0,1\}$-indicator.

Now, for a pair of sites $(x,y)$ where $x$ has no missing values and $y$
has $k_y$ missing values,
\begin{equation} \label{eq:diploidDmissing}
D_2(x,y) = \frac{1}{m-k_y-1} \sum_{i=1}^m c_{x,i}c_{y,i}^{(0)} -
\frac{1}{(m-k_y)(m-k_y-1)} \left( s_x - \sum_{i=1}^m I[c_{y,i}=9]c_{x,i} \right) s_y.
\end{equation}
Indeed, we claim the above equation is actually just a rewriting of
the standard covariance formula \eqref{eq:diploidD}, appropriately
modified now that the covariance is over $m-k_y$ values rather than
$m$:
\begin{itemize}
\item In the sum $\sum_{i=1}^m c_{x,i}c_{y,i}^{(0)}$, missing values
  in $y$ have been 0-replaced, so those terms vanish and the sum
  effectively consists of the desired $m-k_y$ products
  $c_{x,i}c_{y,i}$.
\item Similarly, $s_y$ is equal to the sum of the $m-k_y$ non-missing
  $c_{y,i}$ values.
\item Finally, $s_x - \sum_{i=1}^m I[c_{y,i}=9]c_{x,i}$ represents the
  sum of $c_{x,i}$ over individuals $i$ successfully genotyped at $y$,
  written as the sum $s_x$ over all $m$ individuals minus a
  correction.
\end{itemize}

Substituting \eqref{eq:diploidDmissing} into expression
\eqref{eq:R_x_nomissing} for $R(d)$ and rearranging, we have
\begin{eqnarray*}
R(d) &=& \sumapproxd \frac{I[k_x=0]}{1 + I[k_y=0]} \left(
\frac{1}{m-k_y-1} \sum_{i=1}^m c_{x,i}c_{y,i}^{(0)} \right. \\
& & \phantom{\sum_{\substack{|x-y| \approx d\\k_x = 0}} \frac{1}{1 + I[k_y=0]}} \left. \mbox{}
 - \frac{1}{(m-k_y)(m-k_y-1)} \left( s_x - \sum_{i=1}^m I[c_{y,i}=9]c_{x,i} \right) s_y \right) w_xw_y \\
&=& \sum_{i=1}^m \sumapproxd (I[k_x=0]c_{x,i}w_x) \cdot \left( \frac{1}{1+I[k_y=0]}
\left( c_{y,i}^{(0)} + \frac{I[c_{y,i}=9]s_y}{m-k_y} \right) \frac{w_y}{m-k_y-1} \right) \\
& & \mbox{} - \sumapproxd (I[k_x=0]s_xw_x) \cdot
\left( \frac{s_yw_y}{(1 + I[k_y=0])(m-k_y)(m-k_y-1)} \right).
\end{eqnarray*}
The key point is that we once again have a sum of $m+1$ convolutions
of the form $\sumapproxd f(x)g(y)$ and thus can compute them
efficiently as before.

\section*{One-reference weighted LD}

When computing weighted LD using the admixed population itself as a
reference with one other reference population, a polyache statistic
must be used to obtain an unbiased estimator
(File~\ref{fig:single_anc_unbiased_estimator}).  The form of the
polyache causes complications in our algebraic manipulation; however,
if we restrict our attention to SNPs with no missing data, the
computation can still be broken into convolutions quite naturally,
albeit now requiring $O(m^2)$ FFTs rather than $O(m)$.

As in the two-reference case, the key idea is to split and factorize
the weighted LD formula.  We treat the terms in the polyache
separately and observe that each term takes the form of a constant
factor multiplied by a product of sub-terms of the form $S_{r,s}$,
$p_A(x)$, or $p_A(y)$.  We can use convolution to aggregate the
contributions of such a term if we can factor it as a product of two
pieces, one depending only on $x$ and the other only on $y$.  Doing so
is easy for some terms, namely those that involve only $p_A(x)$,
$p_A(y)$, $S_{r,0}$, and $S_{0,s}$, as the latter two sums depend only
on $x$ and $y$, respectively.

The terms involving $S_{r,s}$ with both $r$ and $s$ nonzero are more
difficult to deal with but can be written as convolutions by further
subdividing them.  In fact, we already encountered $S_{1,1} =
\sum_{i=1}^m c_{x,i}c_{y,i}$ in our two-reference weighted LD
computation: the trick there was to split the sum into its $m$
components, one per admixed individual, each of which could then be
factored into $x$-dependent and $y$-dependent parts and aggregated via
convolution.

Exactly the same decomposition works for all of the polyache terms
except the one involving $S_{1,1}^2$.  For this term, we write
\[
S_{1,1}^2 = \sum_{i=1}^m c_{x,i}c_{y,i} \sum_{j=1}^m c_{x,j}c_{y,j} =
\sum_{i=1}^m \sum_{j=1}^m c_{x,i}c_{x,j} \cdot c_{y,i}c_{y,j},
\]
from which we see that splitting the squared sum into $m^2$ summands
allows us to split the $x$- and $y$-dependence as desired.  The upshot
is that at the expense of $O(m^2)$ FFTs (and restricting our analysis
to SNPs without missing data), we can also accelerate the
one-reference weighted LD computation.

\end{document}